\documentclass[aip,reprint,twocolumn,amsmath,amssymb]{revtex4-1}

\usepackage{graphicx}
\usepackage{bm}

\usepackage[utf8]{inputenc}
\usepackage[T1]{fontenc}
\usepackage[english]{babel}
\usepackage{mathptmx}
\usepackage{xcolor}
\usepackage{stmaryrd}
\usepackage{soul}
\usepackage{amsmath}

\newcommand{\bfrN}{{\bf r}^N}
\newcommand{\bfpN}{{\bf p}^N}
\newcommand{\dbfrN}{{\rm d}{\bf r}^N\ }

\newcommand{\bfq}{{\bf q}}
\newcommand{\bfqT}{{\bf q}^T}
\newcommand{\bfqs}{{\bf q}^*}
\newcommand{\bfqsT}{({\bf q}^*)^T}

\newcommand{\bfPsi}{\boldsymbol{\Psi}}
\newcommand{\DPsi}{\Delta\Psi}

\newcommand{\matA}{\mathbf{A}_0}
\newcommand{\matAtf}{\mathbf{A}_{TF}}
\newcommand{\matAinvtf}{\mathbf{A}^{-1}_{TF}}

\newcommand{\bfA}{\mathbf{A}}
\newcommand{\bfI}{\mathbf{I}}

\newcommand{\bfBrN}{{\bf B}(\bfrN)}
\newcommand{\bfB}{{\bf B}}

\newcommand{\bfE}{{\bf E}}
\newcommand{\bfET}{{\bf E}^T}

\newcommand{\mathH}{\mathcal{H}}
\newcommand{\mathK}{\mathcal{K}}
\newcommand{\mathU}{\mathcal{U}}
\newcommand{\mathUz}{\mathcal{U}_0}
\newcommand{\mathUTF}{\mathcal{U}_{TF}}

\newcommand{\mathZ}{\mathcal{Z}}

\newcommand{\ltf}{l_{TF}}

\newcommand{\avg}[1]{\left\langle\displaystyle #1\right\rangle}

\begin{document}

\title{Microscopic origin of the effect of substrate metallicity on interfacial free energies}

\author{Laura Scalfi}
\affiliation{Physicochimie des \'electrolytes et Nanosyst\`emes Interfaciaux, Sorbonne Universit\'e, CNRS, 4 Place Jussieu F-75005 Paris, France}

\author{Benjamin Rotenberg}
\email[]{benjamin.rotenberg@sorbonne-universite.fr}
\affiliation{Physicochimie des \'electrolytes et Nanosyst\`emes Interfaciaux, Sorbonne Universit\'e, CNRS, 4 Place Jussieu F-75005 Paris, France}
\affiliation{R\'eseau sur le Stockage Electrochimique de l’Energie (RS2E), FR CNRS 3459, 33 Rue Saint Leu 80039 Amiens Cedex, France}

\date{\today}

\begin{abstract}
We investigate the effect of the metallic character of solid substrates on solid-liquid interfacial thermodynamics using molecular simulations. Building on the recent development of a semi-classical Thomas-Fermi model to tune the metallicity in classical molecular dynamics simulations, we introduce a new thermodynamic integration framework to compute the evolution of the interfacial free energy as a function of the Thomas-Fermi screening length. We validate this approach against analytical results for empty capacitors and by comparing the predictions in the presence of an electrolyte with values determined from the contact angle of droplets on the surface. The general expression derived in this work highlights the role of the charge distribution within the metal. We further propose a simple model to interpret the evolution of the interfacial free energy with voltage and Thomas-Fermi length, which allows us to identify the charge correlations within the metal as the microscopic origin of the evolution of the interfacial free energy with the metallic character of the substrate. This new methodology opens the door to the molecular-scale study of the effect of the metallic character of the substrate on confinement-induced transitions in ionic systems, as reported in recent Atomic Force Microscopy and Surface Force Apparatus experiments.
\end{abstract}

\maketitle


The properties of solid-liquid interfaces crucially depend on the delicate balance of interactions of fluid molecules between them and with the surface atoms. For metallic surfaces, the possibility to impose the electric potential provides a handle to control interfacial properties, \emph{e.g.} in electro-wetting (as measured by the contact angle, which reflects interfacial free energies)~\cite{mugele_electrowetting_2005,daub_electrowetting_2007,choudhuri_dynamic_2016}, electro-tunable friction~\cite{sweeney2012a,li_ionic_2014,fajardo2015a,fajardo2015b,pivnic_electrotunable_2020}, or electro-mechanical couplings between surfaces across liquid films~\cite{perez-martinez_surface_2019}. From the theoretical point of view, thermodynamic cycles and continuum electrostatics allow for example to understand the quadratic dependence of interfacial free energies with voltage.

Recent experiments have further demonstrated the dramatic effect of the metallicity of the substrate on the behavior of ionic liquids confined under the tip of an Atomic Force Microscope~\cite{comtet2017a,laine_nanotribology_2020} or in a Surface Force Apparatus~\cite{garcia_nano-mechanics_2017}, even in the absence of voltage. The observed capillary freezing could be rationalized by considering the polarization of the metal by the liquid, usually described in terms of image charges~\cite{netz_debye-huckel_1999,arnold2013a,breitsprecher_electrode_2015,kornyshev2013a,lee2016b}. This polarization depends on the screening of the electric field within the metal and can be accounted for with the Thomas-Fermi model, which quantifies the metallic character by introducing a Thomas-Fermi screening length, $\ltf$. This approach, already successful to investigate the capacitance and structure of electrode-electrolyte interfaces at the mean-field level~\cite{kornyshev_nonlocal_1980,kornyshev_nonlocal_1982}, proved useful to analyze the effect of $\ltf$ on the interactions between ions at the surface~\cite{rochester2013a}, even though analytical calculations of the resulting phase behavior remain out of reach beyond a simplified one-dimensional description~\cite{kaiser2017a}. 

In addition, such continuous approaches neglect molecular features that may play an important role on the interfacial properties, such as the discrete nature of matter (on both the solid and liquid sides), which leads to the layering of the fluid at the surface, the orientation of the solvent molecules and molecular ions, or even possible templating effects modifying the phase behavior of the fluid. Classical molecular simulations have therefore become an essential tool to investigate the interface between fluids and metallic surfaces. The development of methods allowing the simulation in which the voltage between two electrodes can be imposed~\cite{siepmann1995a,reed2007a,limmer2013a,scalfi2020a} opened the way to the molecular understanding \emph{e.g.} of aqueous and non-aqueous electrochemical cells~\cite{willard2009a,merlet2013c,pounds2015a}, of water and ions on platinum surfaces~\cite{limmer2013b,willard2013a,limmer2015b,kattirtzi2017a} or voltage-driven transitions in ionic liquids on graphite electrodes~\cite{merlet2014a,rotenberg2015a} (see \emph{e.g.} Refs.~\citenum{fedorov2014a,scalfi_molecular_2021} for recent reviews).
The role of image charges on the interfacial properties of aqueous solutions~\cite{son_image-charge_2021} and ionic liquids has also been examined in molecular simulations~\cite{merlet2013b,ntim_role_2020}, but the classical description remained limited to the case of ideal conductors, corresponding to a vanishing $\ltf$. For comparison, typical values of the Thomas-Fermi length are 0.5~\AA\ for good metals such as platinum and gold, or 3.6~\AA\ for graphite~\cite{comtet2017a}. We have recently extended the constant-potential method to tune the metallicity in classical molecular simulations based on the Thomas-Fermi model~\cite{scalfi_semiclassical_2020} and an alternative approach to account for screening within the metal with mobile charges has also been proposed in Ref.~\citenum{schlaich2020arXiv}. However direct computation of interfacial free energies for each $\ltf$ remains a great challenge. 

Here, we investigate the influence of metallicity on interfacial free energies using molecular simulations. After a preliminary discussion of contact angles on an insulating or metallic graphite surface, we introduce the relevant thermodynamic quantities to measure the effects of metallicity and voltage. We then present a new thermodynamic integration method to efficiently compute the evolution of interfacial free energies with the Thomas-Fermi screening length $\ltf$ and with voltage. We provide results for empty capacitors and full electrochemical cells consisting of an aqueous electrolyte between graphite or gold electrodes. We finally explain the microscopic origin of the effect of metallicity by analyzing the charge correlations within the metal.

\section{Contact angle on insulating or metallic surfaces}

\begin{figure}[ht!]
\centering
\includegraphics[width=8.7cm]{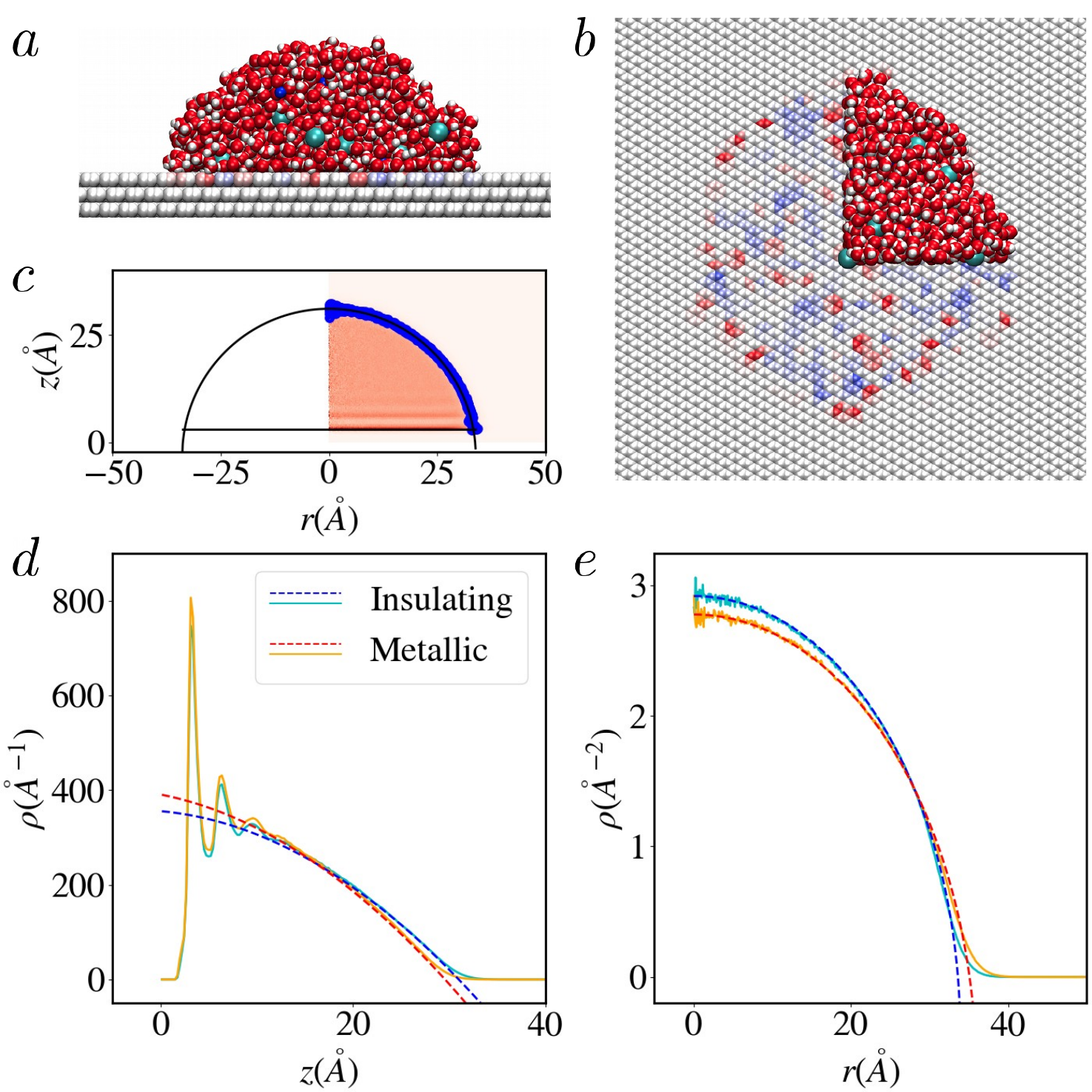}
\caption{Contact angles of aqueous NaCl solution droplets on graphite. Side (a) and top (b) views of the simulated system. In (b), only a quarter of the liquid is shown in order to visualize the instantaneous charges (red for negative and blue for positive) of the carbon atoms in the metallic case. (c) Two-dimensional density map as a function of the height $z$ and the radial dimension $r$; the blue points locate the position of the liquid-vapor interface, fitted to a circle (solid black line) to determine the contact angle. The corresponding one-dimensional density profiles as a function of the height $z$ and radial distance to the droplet center of mass $r$ are shown in panels (d) and (e). In both panels, solid (resp. dashed) lines are the simulation results (resp. predictions for a homogeneous sphere), for the insulating (solid cyan, dashed blue) and metallic (solid orange, dashed red) systems.
}
\label{fig:drop}
\end{figure}

We first illustrate the influence of the metallic character of the substrate on interfacial free energies by considering the contact angle of droplets of a 1M aqueous NaCl solution on graphite. The two limits of insulating or perfectly metallic substrates, corresponding to infinite and vanishing Thomas-Fermi length $\ltf$, respectively, are modeled, all other things being equal, with fixed neutral charges or fluctuating charges determined on-the-fly in order to satisfy a constant-potential condition of the carbon atoms (see Appendix). The system is shown in the metallic case in Figs.~\ref{fig:drop}a and~\ref{fig:drop}b.
The contact angle is determined from the equilibrium shape of the droplet, using a fit of the liquid-vapor interface by a spherical cap (see Fig.~\ref{fig:drop}c). 

In order to facilitate the comparison between the insulating and metallic cases, panels~\ref{fig:drop}d and~\ref{fig:drop}e report the corresponding one-dimensional density profiles as a function of the height $z$ and radial distance to the droplet center of mass $r$. Apart from the layered structure at the interface, the simulation results (solid lines) are very well described by a homogeneous spherical cap (dashed lines -- see Appendix for details). There is a significant (even though small) effect of the metallic character of the substrate on the shape of the droplet. The contact angle is determined from the intersection of the spherical cap with the solid-liquid interface, taken as the plane of the first maximum in the water density profile. This leads to $\theta (\ltf = 0) = 74.6 \pm 0.3^\circ$ and $\theta (\ltf \rightarrow \infty) = 79.6 \pm 0.3 ^\circ$. The metallic surface therefore behaves as more hydrophilic than the insulating one. The link with interfacial free energies is provided by the Young equation
\begin{equation}
\cos \theta(l_{TF}) = \frac{\gamma_{SV}(\ltf) - \gamma_{SL}(\ltf)}{\gamma_{LV}} \,,
\end{equation}
where $\gamma_{SV}$, $\gamma_{SL}$ and $\gamma_{LV}$ are the solid/vapor, solid/liquid and liquid/vapor surface tensions, respectively. By introducing the difference $\Delta \gamma_{SL}(\ltf) = \gamma_{SL}(\ltf)-\gamma_{SL}(0)$ in the solid-liquid interfacial free energy per unit area for a finite $\ltf$ with respect to a perfect conductor ($\ltf=0$), and similarly $\Delta \gamma_{SV}(\ltf)$ for the solid-vapor interface and $\Delta \cos \theta (\ltf)$ for the contact angle, we obtain
\begin{equation}
\label{eq:deltadeltagamma}
\Delta \gamma_{SL}(\ltf) - \Delta \gamma_{SV}(\ltf)= -\gamma_{LV} \Delta \cos \theta(\ltf)
\; .
\end{equation}
The value of $\gamma_{LV}$ is obtained from the normal and tangential components of the pressure tensor, $p_N$ and $p_T$, in a simulation of a slab of electrolyte in vacuum with a box length $L_z$ in the direction perpendicular to the interfaces, as $\gamma_{LV} = (p_N - p_T)L_z/2 = 62.7 \pm 0.6$~mN~m$^{-1}$. Together with $\Delta \cos \theta(\infty) = -0.09 \pm 0.01$, this leads to an excess free energy per unit area $\Delta \gamma_{SL}(\infty)-\Delta \gamma_{SV}(\infty)=5.4\pm0.6$~mN~m$^{-1}$ of the insulating surface with respect to the perfectly metallic one. Note that the change in contact angle, for a given value of the free energy difference (\emph{i.e.} $\Delta \cos \theta$), would be larger if the angle $\theta$ itself were less close to 90$^\circ$.
Another important aspect of metallic surfaces is the possibility to tune the interfacial tension by applying voltage, as discussed below -- opening the way to more dramatic effects such as electrowetting.

\section{Interfacial free energies vs metallicity and voltage}

We now turn to the more direct and systematic study of the effect of metallicity and voltage on interfacial free energies. We consider capacitors consisting of two electrodes characterized by the same $\ltf$ and maintained at fixed voltage $\DPsi$, separated either by vacuum or by an electrolyte (see Fig.~\ref{fig:tfti} below), in order to estimate free energy differences corresponding to solid-vapor ($SV$) or solid-liquid ($SL$) interfaces, respectively. For a capacitor characterized by a voltage-independent capacitance, the accumulated charge at a fixed voltage $\DPsi$ between the electrodes is $Q=C\mathcal{A}\DPsi$, with $C$ the capacitance per unit area and $\mathcal{A}$ the electrode surface area, and the energy stored upon charging is $\Delta U=\frac{Q\DPsi}{2}=\frac{C\mathcal{A}\DPsi^2}{2}$. 
Taking into account the reversible work $-Q\DPsi$ performed on the system, this provides the free energy change per unit area associated with the charge of a given capacitor
\begin{equation}
\label{eq:deltaFSXDpsi}
\frac{F_{SX}^{\DPsi}(\ltf) - F_{SX}^{0}(\ltf)}{\mathcal{A}} =
-\frac{C_{SX}(\ltf)}{2}\DPsi^2 \;,
\end{equation}
with $X=L,V$ and $C_{SX}(\ltf)$ the corresponding capacitance per unit area. One can also consider the difference, for fixed $\Delta\Psi$, between perfectly metallic electrodes and electrodes characterized by a finite $\ltf$
\begin{equation}
\label{eq:deltaFSXltf}
\Delta F_{SX}^{\DPsi}(\ltf) = F_{SX}^{\DPsi}(\ltf) - F_{SX}^{\DPsi}(0) \; .
\end{equation}
Combined with the previous equation, this leads to
\begin{equation}
\label{eq:deltaFSXltf2}
\frac{\Delta F_{SX}^{\DPsi}(\ltf)}{\mathcal{A}}= \frac{\Delta F_{SX}^{0}(\ltf)}{\mathcal{A}}
+ \frac{C_{SX}(0) - C_{SX}(\ltf)}{2}\DPsi^2 \; ,
\end{equation}
(see the thermodynamic cycle in Appendix).
We now introduce
\begin{equation}
\label{eq:deltadeltaFSXltf}
\Delta\Delta F_{SX}^{\DPsi}(\ltf)= \Delta F_{SX}^{\DPsi}(\ltf) - \Delta F_{SV}^{0}(\ltf) \; ,
\end{equation}
which can be efficiently obtained from simulations using the method presented below. Noting that for $\DPsi=0$~V, the two interfaces of each capacitor are identical, this quantity provides the link with the difference in interfacial tensions and contact angle (see Eq.~\ref{eq:deltadeltagamma})
\begin{equation}\label{eq:dftodcostheta}
\Delta\Delta F_{SL}^{0}(\ltf)
= - 2 \mathcal{A}\gamma_{LV} \Delta \cos \theta(\ltf)\; .
\end{equation}

The computation of free energies using molecular simulations requires dedicated approaches, typically involving a thermodynamic path from a reference state with known properties. One possibility is to compute the work of adhesion of the liquid on each surface, as done in Ref.~\citenum{ntim_role_2020} to compare two models of gold in contact with an ionic liquid, in the absence of voltage. Here, we introduce instead a new thermodynamic integration approach, to efficiently compute the free energy change, for a given capacitor and fixed voltage, as a function of the Thomas-Fermi length $\ltf$.

\subsection{Thomas-Fermi thermodynamic integration}

The classical description of metallic electrodes at constant potential using fluctuating charges~\cite{siepmann1995a,reed2007a} was recently extended to materials characterized by a finite $\ltf$~~\citep{scalfi_semiclassical_2020}. In a nutshell, the system involves mobile molecules and ions of the electrolyte, with $N$ point charges at positions $\bfrN$, and $M$ immobile electrode atoms with atom-centered Gaussian charge distributions with fluctuating magnitudes $\bfq$ representing the excess charge. The potential energy $\mathUTF$ can be written as
\begin{align}
\mathUTF(\bfrN, \bfq) = \frac{\bfqT \bfA (\ltf) \bfq}{2}  - \bfqT \bfBrN + \mathcal{C}(\bfrN) \;,  
\end{align}
with $\mathbf{B}$ the vector containing the electrostatic potentials due to the electrolyte on each electrode atom, $\mathcal{C}$ a scalar consisting of all the terms not depending on the electrode charges $\bfq$, and the matrix
\begin{equation}
\bfA (\ltf) = \bfA_0 + \frac{\ltf^2 d}{\epsilon_0} \bfI \, .
\end{equation}
$\bfA_0$ is the electrostatic interaction matrix between electrode atoms for $\ltf=0$, which depends on their positions and the width of the Gaussian distributions and takes into account the periodicity of the system imposed in molecular simulations of condensed matter~\cite{reed2007a,gingrich_ewald_2010,scalfi_molecular_2021}, $d$ is the atomic density of the electrode, $\epsilon_0$ the vacuum permittivity and $\bfI$ is the identity matrix. 

In practice, the set of electrode charges is determined at each time step in order to satisfy the constant potential and the electroneutrality constraints. Such Born-Oppenheimer (BO) dynamics suppress some charge fluctuations from the original constant-potential ensemble, and their effect must be taken into account separately when evaluating some properties such as the differential capacitance~\cite{scalfi2020a}. This is also true for the free energy, obtained from the partition function for a system at fixed $NVT\DPsi$ as $F(\ltf) = - k_BT \ln \mathZ (\ltf)$. We show in Appendix, using thermodynamic integration between the perfect metal case ($\ltf = 0$) and a non-perfect metal with finite $\ltf$ and extending the statistical mechanics derivations of Ref.~\citenum{scalfi2020a} to non-ideal metals, that the free energy difference defined in Eq.~\ref{eq:deltaFSXltf} is given by
\begin{align}
\label{eq:deltaFSXnBOandBO}
\Delta F_{SX}^{\DPsi}(\ltf) &= \Delta F^{nBO}(\ltf) + \int \limits _0 ^{\ltf} \mathrm{d}l \, \frac{ld}{\epsilon_0} \left< \bfqsT \bfqs \right>_{l,\DPsi}  \, ,
\end{align}
where $\Delta F^{nBO}(\ltf)$ is the contribution beyond BO sampling, $\bfqs= \{ q_1, \dots, q_M\}$ are the instantaneous electrode charges in the BO dynamics (for simplicity, we drop the exponent when referring to individual charges), and the ensemble average of the sum of their squares is taken at fixed screening length $l$ and voltage $\DPsi$. 
For the empty capacitor with $\DPsi=0$~V, there is no BO contribution because no charges are induced on the surface. It then follows from Eq.~\ref{eq:deltaFSXnBOandBO} that $\Delta F^{nBO} (\ltf) = \Delta F_{SV}^0(\ltf)$ only depends on the electrode configuration, \emph{i.e.} neither on the presence or absence of electrolyte nor on voltage, and therefore cancels out in differences. In particular, the difference introduced in Eq.~\ref{eq:deltadeltaFSXltf} reduces to the BO contribution to the free energy difference sampled by molecular dynamics:
\begin{align}
\frac{\Delta\Delta F_{SX}^{\DPsi}(\ltf)}{\mathcal{A}} &= \int \limits _0 ^{\ltf} \mathrm{d}l \, \frac{ld}{\epsilon_0 \mathcal{A}} \left< \bfqsT \bfqs \right>_{l,\DPsi}  \, . 
\label{eq::dfboti}
\end{align}
Eq.~\ref{eq::dfboti} provides the practical expression to apply our new thermodynamic integration approach as a function of the metallicity of the electrode. It only requires a set of independent simulations for an ensemble of $\ltf$ values, during which the integrand is straightforwardly obtained from the atomic electrode charges, thereby bypassing the expensive computation of free energies for each $\ltf$.

\subsection{Empty capacitor}

\begin{figure}[ht!]
\centering
\includegraphics[width=8.7cm]{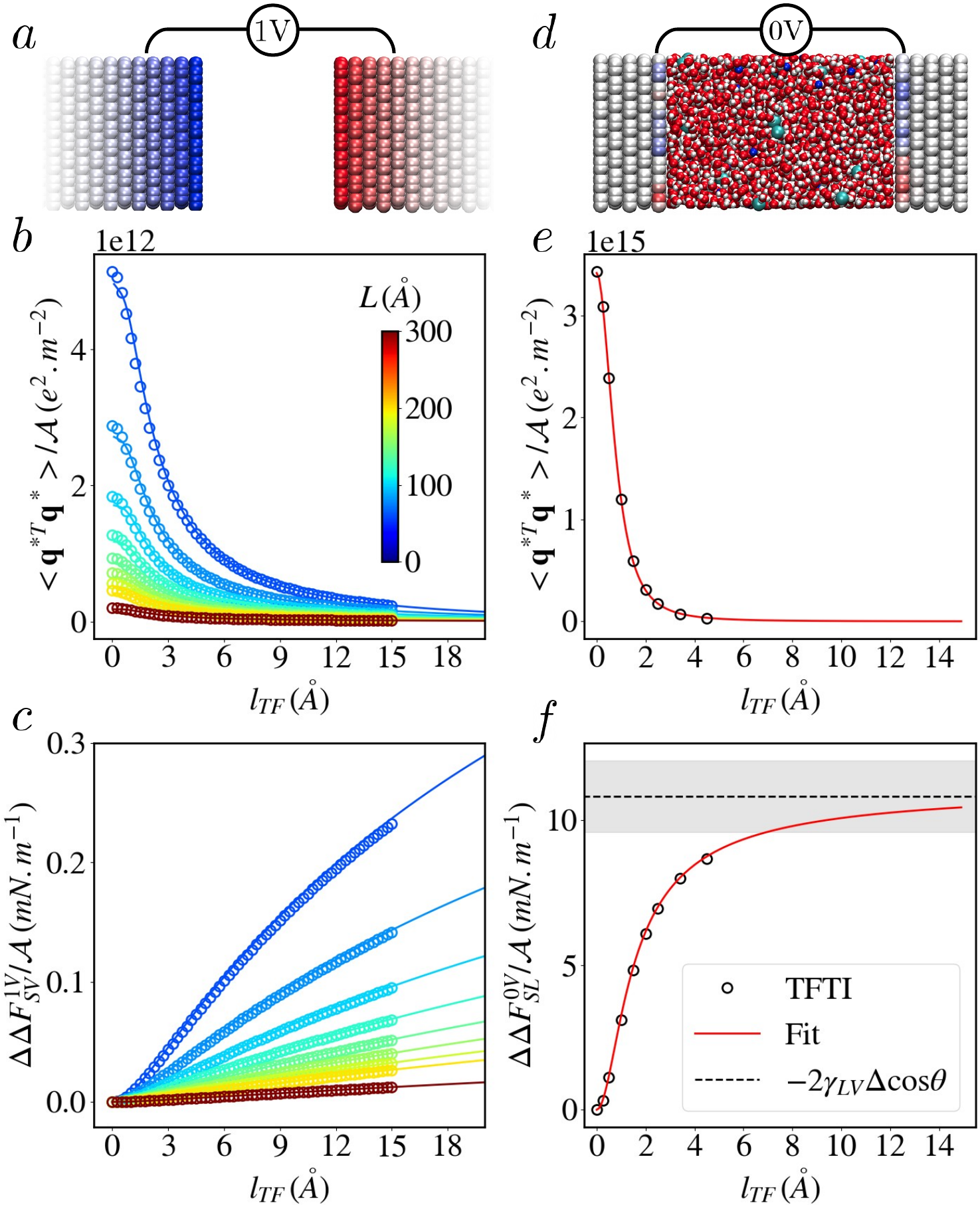}
\caption{Thomas-Fermi Thermodynamic Integration. 
(a) Empty capacitor consisting of two graphite electrodes at $\DPsi = 1$~V, separated by a variable distance $L$ ranging from $60.0$ to $300.0$~\AA\, corresponding to different colors in panels b and c.
(b) Average sum of the square of the atomic electrode charges, per unit area, as a function of $\ltf$.
(c) Free energy difference per unit area $\Delta\Delta F_{SV}^{1V}(\ltf) / \mathcal{A}$ due to a change in the Thomas-Fermi length (see Eq.~\ref{eq:deltadeltaFSXltf}), as a function of $\ltf$, computed from Eq.~\ref{eq::dfboti}.
(d) Capacitor consisting of two graphite electrodes at $\DPsi = 0$~V, separated by a distance $L=56.2$~\AA, and a 1M NaCl aqueous solution.
(e) Same as panel b for the aqueous capacitor.
(f) Free energy difference per unit area $\Delta\Delta F_{SL}^{0V}(\ltf)/ \mathcal{A}$ and asymptotic prediction from contact angles using Eq.~\ref{eq:dftodcostheta} (for the difference between metallic and insulator, with $\ltf=0$ and $\ltf\to\infty$, respectively). 
In panels b-c and e-f, open circles are simulation data, while solid lines are the analytical expression Eq.~\ref{eq::dfboempty} for panel c (and corresponding for b) and a fit of the form Eq.~\ref{eq::dfbofull0Vfit} for panel f (and corresponding for e), with parameters adjusted on the simulation data of panel e.
}
\label{fig:tfti}
\end{figure}

We first consider the case of an empty capacitor, consisting of two graphite electrodes of surface area $\mathcal{A}$, each with $n = 50$ layers with interplane distance $a$, separated by a range of distances $L$ between the first atomic planes (see Fig.~\ref{fig:tfti}a), held at a potential difference $\Delta \Psi = 1$~V. In such a system, the total charge on each electrode is well predicted by continuum theories, the charge per plane decreases exponentially within the electrode (with a decay length $\ltf$) and is homogeneous within each atomic plane~\cite{scalfi_semiclassical_2020}. By further assuming that the number of planes is large, one can write the charge per plane (indexed by $k\in [\![ 1,\infty [\![$ in each electrode) as $Q_k=\pm Q e^{-(k-1)a/\ltf}(1-e^{-a/\ltf})$. Introducing $m$ the number of atoms per plane, related to the atomic density of the electrode by $d=m/a\mathcal{A}$, and the capacitance $C_{SV}(\ltf)=\epsilon_0/(L+2\ltf)$ corresponding to a vacuum slab and two Thomas-Fermi electrodes in series, Eq.~\ref{eq::dfboti} finally leads to
\begin{equation}
\frac{\Delta\Delta F_{SV}^{\DPsi}(\ltf)}{\mathcal{A}} = 2\epsilon_0 \Delta\Psi^2 \int \limits _0 ^{\ltf} \mathrm{d}l \, \frac{1}{(L + 2 l)^2} f\left(\frac{a}{l}\right) \, ,
\label{eq::dfboempty}
\end{equation}
where $f(x)=\frac{1}{x}\sum_{k=1}^\infty e^{-2(k-1)x}(1-e^{-x})^2 = \frac{(1-e^{-x})^2}{x(1-e^{-2x})}$ comes from summing over the electrode planes.
In the continuum limit ($a\to0$), this expression simplifies to $\Delta\Delta F_{SV}^{\DPsi}(\ltf)\approx  \epsilon_0 \mathcal{A}\ltf\Delta\Psi^2/L(L + 2 \ltf)$, consistently with Eqs.~\ref{eq:deltaFSXltf2}-\ref{eq:deltadeltaFSXltf}. For insulating materials ($\ltf\to\infty$), the difference with respect perfect conductors further reduces to $\Delta\Delta F_{SV}^{\DPsi}(\infty)\approx\epsilon_0 \mathcal{A}\Delta\Psi^2/2L$.

Results from the calculations with explicit electrode atoms are given in Fig.~\ref{fig:tfti}b-c for the average $\left< \bfqsT \bfqs \right> / \mathcal{A}$ and the free energy difference per unit area as a function of the Thomas-Fermi length. The evolution of $\Delta\Delta F_{SV}^{\DPsi}(\ltf)$ as a function of both $\ltf$ and $L$ is very well described by Eq.~\ref{eq::dfboempty}. We have also checked the quadratic dependence of the result with the voltage $\DPsi$ (values for $\DPsi = 2$~V are given in Fig.~\ref{fig:tfti-2V} in Appendix). We finally note that in the absence of an electrolyte, for this typical voltage of 1~V the order of magnitude of the change in free energy per unit area due to a change in metallicity is small compared \emph{e.g.} to the liquid-vapor surface tension of water $\gamma_{LV}$ or even to $\Delta \gamma_{SL}(\infty)-\Delta \gamma_{SV}(\infty)$ deduced from the contact angles (see Eq.~\ref{eq:deltadeltagamma}).

\subsection{Electrochemical cell}

We now move to the more complex case of an electrochemical cell consisting of the same graphite electrodes (with only $n=5$ planes, which reduces the accessible range of $\ltf$ values) separated by a distance $L$, with a 1M NaCl aqueous solution as electrolyte, in the absence of voltage ($\DPsi=0$~V). This system is shown in Fig.~\ref{fig:tfti}d, while the results for the evolution of $\left< \bfqsT \bfqs \right>/\mathcal{A}$ and of $\Delta\Delta F_{SL}^{0V}/\mathcal{A}$ as a function of $\ltf$ are shown in panels~\ref{fig:tfti}e-f. $\Delta\Delta F_{SL}^{0V}$ increases with $\ltf$, with a shape similar to but different from $\Delta\Delta F_{SV}^{1V}$, and is more than an order of magnitude larger than the latter. The solid line in panel~\ref{fig:tfti}f corresponds to an empirical fit of the form (see also Eq.~\ref{eq::dfbofull0V} below)
\begin{equation}
\label{eq::dfbofull0Vfit}
\frac{\Delta\Delta F_{SL}^{0V}(\ltf)}{\mathcal{A}} = 2 k_BT \int \limits _0 ^{\ltf} \mathrm{d}l \, \frac{1}{\gamma_0 + \gamma_2 l^2 } f\left(\frac{a}{l}\right) \, ,
\end{equation}
with $f$ defined below Eq.~\ref{eq::dfboempty}, and
where $\gamma_0$ and $\gamma_2$ are two parameters adjusted on $\avg{\bfqsT\bfqs}/\mathcal{A}$ (panel~\ref{fig:tfti}e). At this stage, we will only emphasize that its extrapolation for $\ltf\to\infty$ is consistent with the value obtained from the difference in contact angles, also indicated as a dashed line in panel~\ref{fig:tfti}f. This further validates our new thermodynamic integration approach to compute free energy differences as a function of $\ltf$. In addition, the transition from metallic to insulating behavior occurs mainly for $\ltf$ in the range of a few \AA.

\subsection{Effect of voltage on interfacial free energies}

\begin{figure}[ht!]
\centering
\includegraphics[width=8.7cm]{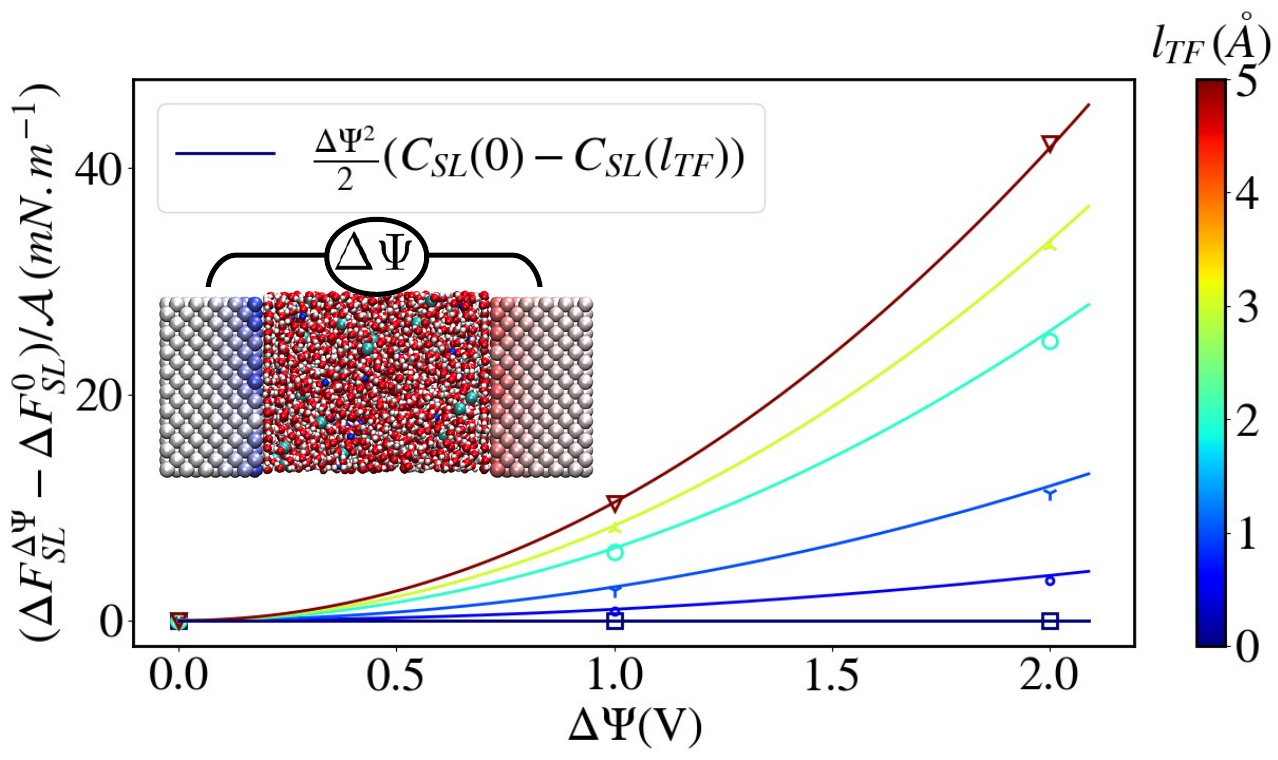}
\caption{
Effect of voltage $\DPsi$. Free energy change
$\Delta F_{SL}^{\DPsi}(\ltf)$ with respect to the case of a perfect conductor (see Eq.~\ref{eq:deltaFSXltf}) as a function of $\DPsi$ in the case of an electrochemical cell consisting of a 1M NaCl aqueous electrolyte between gold electrodes (see inset). Results are shown per unit area, after subtraction of the value in the absence of voltage, $\Delta F_{SL}^{0}(\ltf)$. The simulation results (symbols) are shown for several values of the Thomas-Fermi length $\ltf$, indicated by their color, and compared to the quadratic prediction Eq.~\ref{eq:deltaFSXltf2} (lines) using the capacitances per unit area determined from the average charge.
}
\label{fig:dpsi_dep}
\end{figure}

Fig.~\ref{fig:dpsi_dep} illustrates the effect of voltage $\DPsi$ on the interfacial free energy, for an electrochemical cell consisting of two gold electrodes with their (100) face in contact with a 1M NaCl aqueous solution. The free energy difference $\Delta F_{SL}^{\DPsi}(\ltf)$ with respect to the case of a perfect conductor (see Eq.~\ref{eq:deltaFSXltf}) follows the expected quadradic dependence on voltage and is in perfect agreement with the prediction Eq.~\ref{eq:deltaFSXltf2}, using the capacitances $C(0)$ and $C(\ltf)$ obtained from the average charge of the electrodes (the same capacitance is obtained at 1 and 2~V).

\section{Microscopic origin: the role of lateral correlations}

Overall, the above results demonstrate the relevance of molecular simulations combined with our new thermodynamic integration as a function of $\ltf$ in order to investigate the effect of metallicity and voltage on interfacial free energies. Importantly, Eq.~\ref{eq::dfboti}, derived from statistical-physical considerations, emphasizes the role of the atomic charge distribution within the metal. In the following, we further exploit the possibilities offered by molecular simulations to investigate the microscopic origin of this charge distribution and the effect of metallicity on the latter. 

\subsection{Metallicity changes lateral charge correlations in the solid}
\begin{figure}[htb!]
\centering
\includegraphics[width=8.7cm]{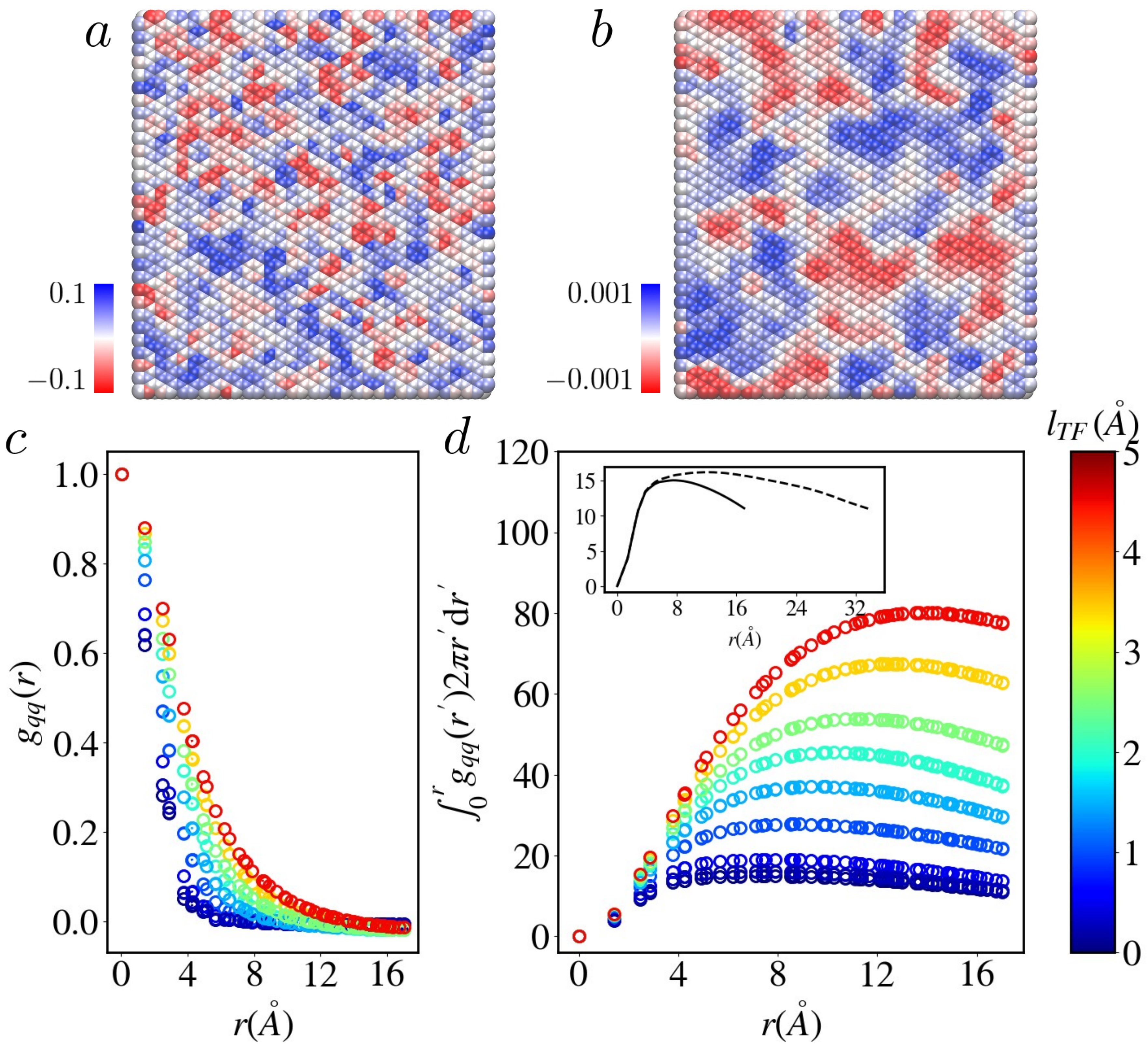}
\caption{
Effect of metallicity on lateral charge correlations. Instantaneous charge distribution (color bar in units of the elementary charge $e$) on the first electrode plane for $\ltf = 0.0$~\AA\ (a) and $\ltf = 4.5$~\AA\ (b), for a 1M NaCl / graphite capacitor at $\DPsi=0$~V. Charge-charge radial distribution function $g_{qq}(r)$ in the first atomic plane (c) and its integral $\int _0 ^r g_{qq}(r') 2 \pi r' {\rm d}r'$ (d) for a range of $\ltf$ values. The inset of panel (d) shows the integral of $g_{qq}$ for $\ltf=0.0$ for the main simulation box (solid line) and for one with double size in the $x$ and $y$ directions (dashed line, corresponding to the snapshots a and b).
}
\label{fig:correlations}
\end{figure}

The heterogeneity of the charge distribution within the metal reflects both the structure of the interfacial fluid and how the metal is polarized by each source. 
While the structure of water in the first adsorbed layer hardly depends on the screening length $\ltf$ (see Figs.~\ref{fig:gOO} and \ref{fig:worient} in Appendix), the latter has a large influence on the charge distribution, as illustrated for the first electrode plane in Figs.~\ref{fig:correlations}a and~\ref{fig:correlations}b, for the previous graphite-aqueous NaCl systems (with a four times larger surface area), at $\DPsi = 0V$. Fig.~\ref{fig:correlations}c then shows the corresponding in-plane charge-charge correlation function $g_{qq}(r)=\left< \delta q(r) \delta q(0) \right>/\left< \delta q^2 \right>$, with $\delta q= q - \left<q\right>$ the local deviation from the average charge in the first electrode plane. The decay of $g_{qq}(r)$ is slower for larger $\ltf$, consistently with the decay of the charge distribution inside the metal induced by an external point charge~\cite{scalfi_semiclassical_2020}. In order to quantify the extent of these lateral correlations, we introduce the correlation surface
\begin{equation}
\label{eq:Scorrdef}
S_{\rm corr}=  \int_{0}^{\infty} g_{qq}(r) 2\pi r{\rm d}r  \; , 
\end{equation}
which will be discussed below. Similar quantities can be defined for each electrode plane $k$. For a homogeneous charge distribution (as in empty capacitors), $S_{\rm corr}\to\infty$, while for a completely random charge distribution, $S_{\rm corr}=S_1 = \mathcal{A}/m$, the area per atom. In practice, for a perfect metal ($\ltf=0$), the correlation length corresponding to $S_{\rm corr}$ reflects that of the interfacial fluid~\cite{limmer2013a}. Fig.~\ref{fig:correlations}d shows the running integral corresponding to Eq.~\ref{eq:Scorrdef}. Even though the finite box size does not allow to reach a plateau for large $r$, the inset shows that the value of the running integral at half the box size is roughly independent of the latter and the corresponding value will be used in the discussion below.

\subsection{From charge correlations to interfacial free energies}

We now use the above microscopic information on the charge distribution to derive a simple model of the surface free energy to understand its evolution with $\ltf$ and with voltage. Starting from Eq.~\ref{eq::dfboti}, we group the sum of the square of atomic charges $\left< \bfqsT \bfqs \right>$ by planes and express, within each plane $k$, the contribution of each atom to the sum as $\left< q_k^2 \right> =\left< q_k \right>^2 + \left< \delta q_k^2 \right>$. The first term is simply $\left< Q_{k} \right>^2/m^2$, with the average charge of the plane and the number of atoms per plane. For the second, we introduce $\alpha_k=m\left< \delta q_k^2 \right>/\left< \delta Q_{k}^2\right>$ to relate the atomic charge fluctuations to that of the plane. 
We then consider the charge $Q_\mathcal{S} = \frac{m}{\mathcal{A}}\iint_{\mathcal{S}} {\rm d}\mathcal{S}\, \delta q({\bf r})$ of a surface element $\mathcal{S}$ much larger than the correlation length and approximate (see Ref.~\citenum{limmer2013a})
\begin{align}
\left<(\delta Q_\mathcal{S})^2\right> 
& =  \left(\frac{m}{\mathcal{A}}\right)^2
\iint_{\mathcal{S}} {\rm d}\mathcal{S}'\iint_{\mathcal{S}} {\rm d}\mathcal{S}\,  \left< \delta q({\bf r}) \delta q({\bf r}') \right> \nonumber \\
& \approx  \left(\frac{m}{\mathcal{A}}\right)^2 \mathcal{S} \left< \delta q^2 \right> \int_{0}^{\infty} g_{qq}(r) 2\pi r {\rm d}r
\; ,
\end{align}
by introducing the relative position $r$ in polar coordinates and extending the integral to infinity. Applying this relation to a whole electrode plane with area $\mathcal{A}$ and introducing the correlation surface Eq.~\ref{eq:Scorrdef}, we obtain $\alpha_k=S_1/S_{\rm corr,k}$.
Summing over the $m$ atoms in each plane, and over the planes of both electrodes leads to
$\left< \bfqsT \bfqs \right> \approx  2 \sum \limits_{k=1}^{\infty} \left[ \frac{\left< Q_{k} \right>^2}{m} + \frac{\left< \delta Q_{k}^2 \right>S_1}{S_{{\rm corr},k}} \right]$.
We then make two assumptions, discussed below, (i) that the charge decays exponentially within the electrode, to relate the charge of each plane to the total charge $Q$ of the electrode, and (ii) that the correlation surface is the same in all planes. Performing the sums over planes as for the empty capacitor, we obtain
\begin{align}
\label{eq:qsTqs1}
\frac{\left< \bfqsT \bfqs \right>}{\mathcal{A}} &\approx \frac{2S_1 a}{\ltf}f\left(\frac{a}{\ltf}\right)
\left[ \left( C_{\rm int}\Delta\Psi\right)^2 + \frac{k_BT C_{\rm diff}^{BO} }{S_{\rm corr}} \right]
\end{align}
using the function $f$ defined below Eq.~\ref{eq::dfboempty} and that the total charge can be expressed using the integral capacitance per unit area $Q=C_{\rm int}\mathcal{A}\Delta\Psi$, while its fluctuations provide the electrolyte (Born-Oppenheimer) contribution to the differential capacitance per unit area~\cite{scalfi2020a} as $C_{\rm diff}^{BO}=\left< \delta Q^2 \right>/(k_BT \mathcal{A})$. Introducing the above results in Eq.~\ref{eq::dfboti} finally yields
\begin{equation}
\label{eq::dfbofull}
\Delta\Delta F_{SL}^{\DPsi}(\ltf)
= \Delta\Delta F_{SL}^{0V}(\ltf)
+ \mathcal{B} \DPsi^2
\end{equation}
where
\begin{equation}
\label{eq::dfbofull0V}
\frac{\Delta\Delta F_{SL}^{0V}(\ltf)}{\mathcal{A}} = \frac{2 k_BT}{\epsilon_0} \int \limits _0 ^{\ltf} \mathrm{d}l \, \frac{C_{\rm diff}^{BO}(l)}{S_{\rm corr}(l)} f\left(\frac{a}{l}\right) \, ,
\end{equation}
and $\mathcal{B}  = \frac{2\mathcal{A}}{\epsilon_0} \int \limits _0 ^{\ltf} \mathrm{d}l \, C_{\rm int}(l)^2 f\left(\frac{a}{l}\right) \xrightarrow[a\to 0]{} \frac{\mathcal{A}}{\epsilon_0}\int \limits _0 ^{\ltf} \mathrm{d}l \, C_{\rm int}(l)^2$. 
In the continuum limit ($a\to0$), we find that Eq.~\ref{eq::dfbofull} reduces to Eqs.~\ref{eq:deltaFSXltf}, \ref{eq:deltaFSXltf2} and \ref{eq:deltadeltaFSXltf} when $1/C_{\rm int}(\ltf)=1/C_{\rm int}(0)+2\ltf/\epsilon_0$, which corresponds to an ideal capacitor in series with two Thomas-Fermi electrodes.

\begin{figure}
\centering
\includegraphics[width=8.7cm]{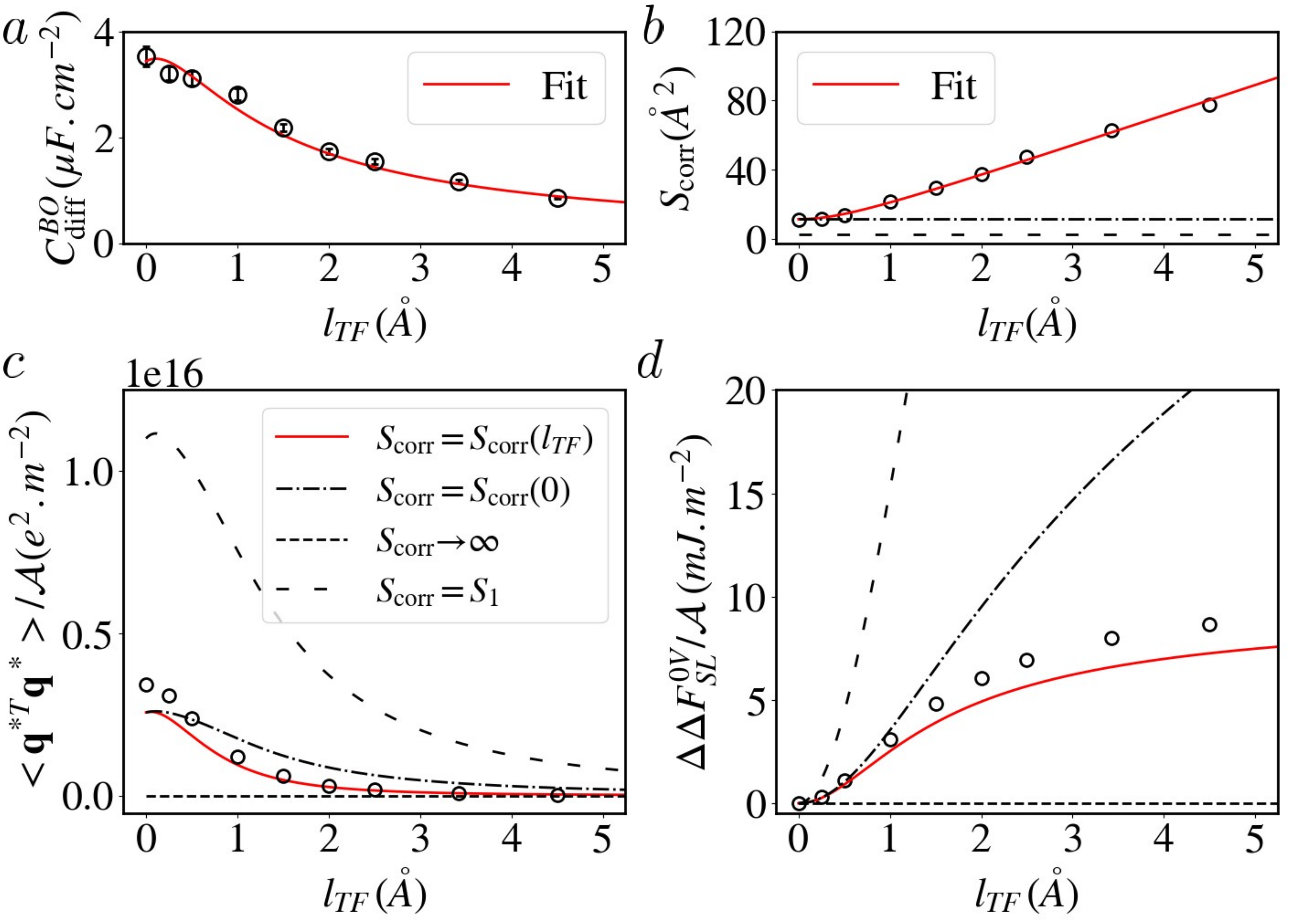}
\caption{From charge correlations to interfacial free energies.
(a) Differential capacitance $C_{\rm diff}^{BO}$ as a function of the Thomas-Fermi length $\ltf$ computed from the electrode charge fluctuations at $\DPsi = 0.0$~V, for a 1M NaCl / graphite capacitor (open black circles); the line is a fit (see text) used in panels (c) and (d). 
(b) Correlation surface $S_{\rm corr}$ obtained from the charge-charge correlation function (see Eq.~\ref{eq:Scorrdef}); the solid red line is a fit (see text) used in panels (c) and (d), while the dashed black line indicates the area per atom $S_1 = \mathcal{A}/m$ and the dash-dotted black line the constant value $S_{\rm corr}(\ltf=0)$. Average squared charges per unit area $\langle \bfqsT \bfqs \rangle/\mathcal{A}$ (c) and interfacial free energy difference per unit area $\Delta\Delta F_{SL}^{0V}(\ltf)/\mathcal{A}$ (d): simulation results (symbols) are compared to the predictions of Eqs.~\ref{eq:qsTqs1} and~\ref{eq::dfbofull0V} (solid red line) using the fits of panels (a) and (b) or neglecting the dependence of $S_{\rm corr}$ with $\ltf$, \emph{i.e.} assuming perfectly metallic ($S_{\rm corr}(\ltf)=S_{\rm corr}(0)$, dash-dotted line), homogeneous ($S_{\rm corr}\to\infty$, dotted line) or uncorrelated ($S_{\rm corr}=S_1$, dashed line) charge distributions.
}
\label{fig:model}
\end{figure}

Beyond the potential-dependent part, the present analysis highlights the deep connection between the interfacial free energy and the charge correlations within the metal, as evident from the presence of the correlation surface $S_{\rm corr}$ in Eq.~\ref{eq::dfbofull0V}. This simple model only requires the evolution of the differential capacitance $C_{\rm diff}^{BO}$ and $S_{\rm corr}$ with $\ltf$ to compute $\Delta\Delta F_{SL}^{0V}(\ltf)/\mathcal{A}$. Results from simulations at $\DPsi=0$~V are shown in Figs.~\ref{fig:model}a and~\ref{fig:model}b, respectively, together with empirical fits of the form $C_{\rm diff}^{BO}=\epsilon_0/(\gamma'_0+2\ltf+\gamma'_1/(\gamma'_2+\ltf))$ and $S_{\rm corr}=\sqrt{\gamma_0''+\gamma_2''\ltf^2}$ for further use. When the Thomas-Fermi length increases, the capacitance decreases due the delocalization of the charge deeper inside the surface which screens the potential (consistently with the equivalent circuit picture of capacitors in series); this delocalization also manifests laterally with an increase in the correlation area. 

Fig.~\ref{fig:model}c then compares $\left< \bfqsT \bfqs \right>/\mathcal{A}$ measured in simulations to the approximation Eq.~\ref{eq:qsTqs1} using the above-mentioned fits of $C_{\rm diff}^{BO}$ and $S_{\rm corr}$. Also shown are the predictions of this approximation when neglecting the dependence of $S_{\rm corr}$ with $\ltf$, \emph{i.e.} assuming perfectly metallic ($S_{\rm corr}(\ltf)=S_{\rm corr}(0)$), homogeneous ($S_{\rm corr}\to\infty$) or uncorrelated ($S_{\rm corr}=S_1$, the surface per atom) charge distributions. Even though some deviations are observed for small $\ltf$, the agreement is remarkable when considering $S_{\rm corr}(\ltf)$, unlike either of the two extreme distributions. Fig.~\ref{fig:model}d finally shows the same comparison for $\Delta\Delta F_{SL}^{0V}(\ltf)$, now with Eq.~\ref{eq::dfbofull0V}. The deviations are now more evident for large $\ltf$, due to the accumulation of errors upon integration, but the semi-quantitative agreement obtained when considering $S_{\rm corr}(\ltf)$, compared to the dramatic failure of both the random and homogeneous distributions and even assuming that the correlations are the same as in the perfect metal case, confirms the crucial role of charge correlations within the metal on the evolution of the interfacial free energy with $\ltf$.

As noted above, the transition from metallic to insulating behavior of the interfacial free energies occurs mainly for $\ltf$ in the range of a few \AA. Translating the correlation area into an effective correlation length $l_{\rm corr}(\ltf)=\sqrt{S_{\rm corr}(\ltf)/\pi}$ (which corresponds to approximating $g_{qq}(r)$ by a Heaviside function), we find that this range corresponds to a change from $l_{\rm corr}\approx1.9$~\AA\ for $\ltf=0$ to $l_{\rm corr}\approx5.0$~\AA\ for $\ltf=4.5$~\AA. We can compare this to the position of the first maximum of the 2D radial distribution function of water molecules in contact with the surface, $r_{OO}^{2D}\approx2.7$~\AA\ (see Fig.~\ref{fig:gOO} in Appendix). This supports the conclusion that the transition for the free energy corresponds to a change from a regime where the charge correlations reflect the structure of the liquid to another where the poor screening within the metal gradually homogenizes the charge distribution.

\subsection{Discussion}
\begin{figure}[ht!]
\centering
\includegraphics[width=8.7cm]{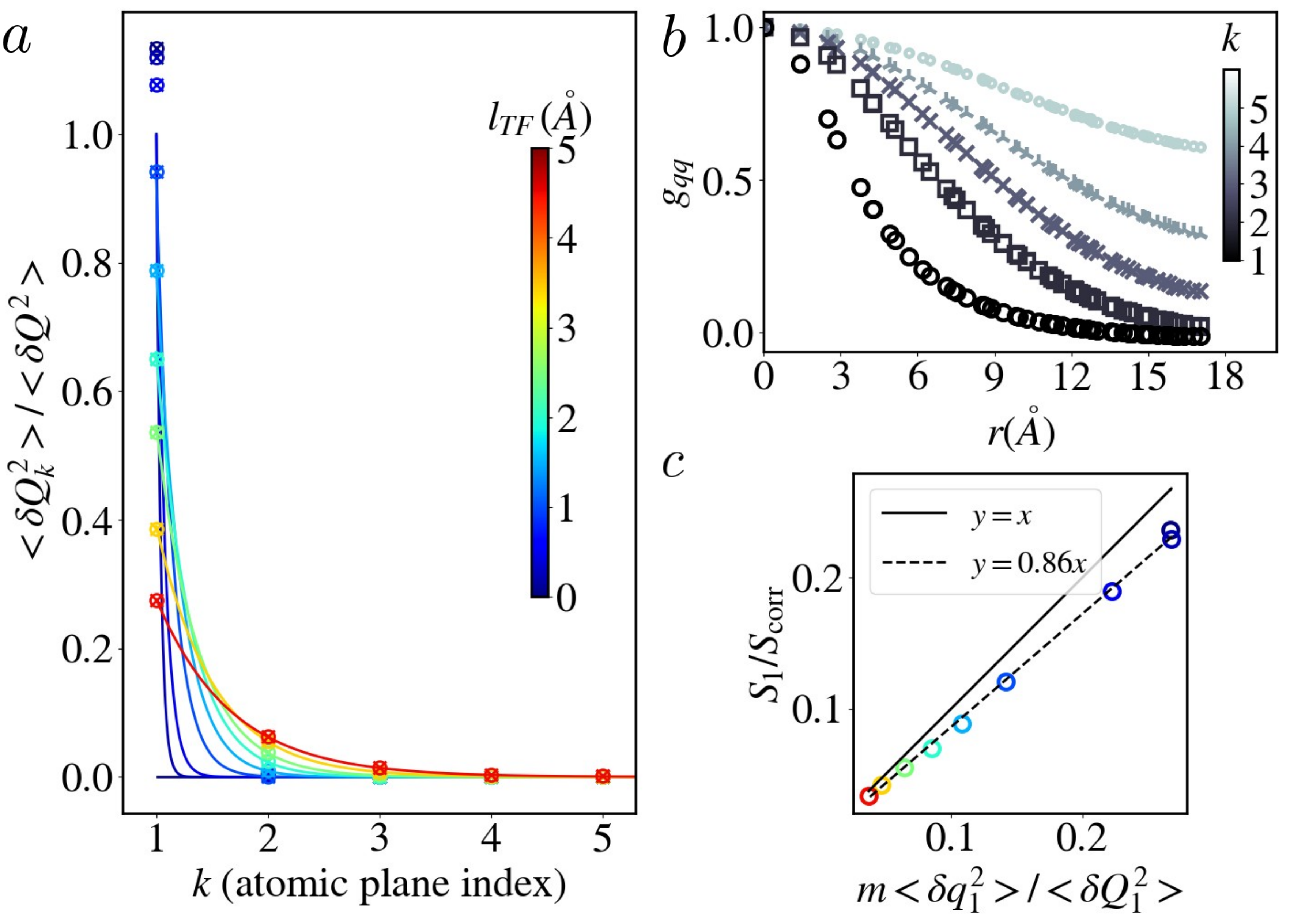}
\caption{
Testing the assumptions of the model for a 1M NaCl / graphite capacitor at $\DPsi = 0.0$~V.
(a) Ratio between the variance of the charge of each plane $k$ and that of the total charge of the electrode, $\left<\delta Q_k^2\right>/\left<\delta Q^2\right>$, for a range of $\ltf$ values from $0.0$ to $4.5$~\AA\, indicated by the colorbar; the lines are analytical predictions for an exponential decay, with $a$ the interplane distance and a decay length $\ltf/2$.
(b) In-plane charge-charge radial distribution function $g_{qq}$ as a function of the distance $r$ for $\ltf=4.5$~\AA\, for the different planes of the electrodes, where $k$ is the atomic plane index from 1, in contact with the electrolyte (black circles) to 5 (light grey small circles).
(c) Ratio $S_1/S_{\rm corr}$ as a function of $m\avg{\delta q_1^2}/\avg{\delta Q_1^2}$ in the first atomic plane, for a range of $\ltf$ shown in the colorbar. The solid black line corresponds to $y=x$ while the dashed line is a fit of the form $y=\gamma x$.
}
\label{fig:assumptions}
\end{figure}

In order to explain the remaining difference between the exact result Eq.~\ref{eq::dfboti} and the approximation Eq.~\ref{eq::dfbofull0V}, we examine the assumptions that allowed us to obtain this simple expression. Beyond the convergence of the integral defining $S_{\rm corr}$ (Eq.~\ref{eq:Scorrdef}) for a finite-size system, discussed above, and the possible errors due to the numerical integration in Eq.~\ref{eq::dfboti} that can be reduced systematically by increasing the number of simulated $\ltf$ values, we have assumed that the charge decays exponentially within the electrode, with a decay length $\ltf$. Such an assumption is consistent, for the average charge in each plane, with the observation that for a finite voltage the average potential decays exponentially within the electrodes (as shown for gold electrodes in Ref.~\citenum{scalfi_semiclassical_2020}). For the charge fluctuations also considered in the present model, we test it for $\DPsi = 0V$ by considering $\langle\delta Q_k^2\rangle/\langle\delta Q^2\rangle$, reported for a range of $\ltf$ values and a 1M NaCl / graphite capacitor in Fig.~\ref{fig:assumptions}a. The exponential assumption is satisfied, except for very small $\ltf$ (smaller than the interplane distance $a$), and the corresponding decay length is $\ltf/2$, as expected.

We further assumed that the correlation surface $S_{\rm corr,k}$ is the same in all electrode planes and used the value $S_{\rm corr}$ measured in the first plane in contact with the liquid. Fig.~\ref{fig:assumptions}b shows the charge-charge correlation function in the five planes for $\ltf=4.5$~\AA\ and clearly demonstrates that this assumption is not accurate. Its shortcomings are however mitigated by the exponential decay of the contribution of each plane. Finally, we introduced the correlation surface to link the atomic charge fluctuations to those of each plane, which required assuming in particular that the area of the electrode, $\mathcal{A}\approx1250$~\AA$^2$, is much larger than $S_{\rm corr}$: Fig.~\ref{fig:model}b confirms that this is indeed the case. 
The ratio $m\avg{\delta q_1^2}/\avg{\delta Q_1^2}$ can also be estimated directly and is compared to $S_1/S_{\rm corr}$ in Fig.~\ref{fig:assumptions}c. While the agreement is not quantitative, the linear correlation with a slope $\approx1.2$ clearly supports our microscopic interpretation of the effect of metallicity on the interfacial free energy, via the lateral charge correlations within the metal. 

\section{Conclusion}

A new thermodynamic integration method as a function of the Thomas-Fermi length allowed us to investigate the effect of the metallic character of solid substrates on interfacial free energies using molecular simulations. This approach was validated against analytical results for empty capacitors and by comparing the predictions in the presence of an electrolyte with values determined from the contact angle of droplets on the surface. The general expression derived in this work highlights the role of the charge distribution within the metal. We proposed a simple semi-analytical model to interpret the evolution of the interfacial free energy with voltage and Thomas-Fermi length, which allowed us to identify the charge correlations within the metal as the microscopic origin of the evolution of the interfacial free energy with the metallic character of the substrate. 
This new methodology further opens the door to the molecular-scale study of the effect of the metallic character of the substrate on  confinement-induced transitions in ionic systems, as reported in AFM and SFA experiments, with the possibility to investigate the relative stability of interfacial crystals with respect to their melt. This will in particular allow to go beyond the analytical calculations on the effect of Thomas-Fermi screening for a 1D ionic crystal near a substrate~\cite{kaiser2017a} to the full three-dimensional interface, both in solid and liquid phases, and to analyze the effect of lateral correlations. Since the charge distribution is known to have a large impact on solid-liquid friction~\cite{xie_liquid-solid_2020}, one could also consider the possibility to extend the present approach to examine the effect of metallicity on dynamical properties.


\section*{\small Materials and methods}
{\small All simulation details, including the description of the systems, the computation of interactions and the constant-potential ensemble are available in the Appendix.
}

\section*{\small Data availability} 
{\small MetalWalls~\cite{marin-lafleche_metalwalls_2020}, the molecular dynamics code used for this study, is available open-source at \url{ https://gitlab.com/ampere2/metalwalls}. The data that supports the findings of this study is available upon reasonable request to the authors.}

\begin{acknowledgments}
The authors thank Mathieu Salanne, Lyd\'eric Bocquet, Beno\^it Coasne and David Limmer for useful discussions. This project has received funding from the European Research Council  under the European Union's Horizon 2020 research and innovation programme (grant agreement No. 863473). This work was supported by the French National Research Agency (Labex STORE-EX, Grant  ANR-10-LABX-0076, and project NEPTUNE, Grant  ANR-17-CE09-0046-02). The authors acknowledge HPC resources granted by GENCI (resources of CINES, Grant No A0070911054).
\end{acknowledgments}

\appendix

\section{Contact angles}
Contact angles were determined from density maps in the $(r,z)$ plane (see Fig.~1c of the main text), with $r$ the radial distance to the center of mass of the droplet and $z$ the height with respect to the first graphite layer. The liquid-vapor interface is located from the points where the average local density is equal to half of the bulk density $\rho_0$ at the center of the droplet (in blue on Fig.~1c of the main text). This set of points is then fitted using a circle of radius $R_C$ and centered in ($r=0, z=z_C$) and the value of the contact angle $\theta$ is obtained by the intersection of this fit with the interface plane, taken as the position of the first water layer $z_W=3.12$~\AA. This choice has a small influence on the contact angle values, which is within the errorbars and does not modifiy the conclusions of the study. 

For a homogeneous sphere of density $\rho_0$, radius $R_C$ and centered in ($r=0, z=z_C$), cut by a plane in $z=z_0$ (taken empirically at $z_W/2$), the corresponding one-dimensional densities are given by
\begin{align}
\rho(z) &= \rho_0 \pi (R_C^2 - (z-z_C)^2) & \label{eq:rhoz_sph}\\
\rho(r) &= \rho_0 \left(z_C - z_0 + \sqrt{R_C^2 - r^2}\right) \,, \label{eq:rhor_sph}
\end{align}
shown along with the simulation results in Fig.~1d-e of the main text.

\section{Derivation of the Thomas-Fermi Thermodynamic Integration method}
We consider a system of $N$ mobile atoms of electrolyte, with positions $\bfrN$ and momenta $\bfpN$, and $M$ electrode atoms that carry a Gaussian charge distribution with fluctuating magnitude $\bfq = \{q_1, q_2, \dots, q_M\}$, in a finite volume $V$, at a temperature $T$ and a fixed voltage $\DPsi$ between the electrodes. The free energy associated with a change in Thomas-Fermi (TF) screening length $\ltf$ is computed starting from the definition of free energy $F^{\DPsi} = - \beta^{-1} \ln  \mathZ$, with $\mathZ$ the partition function corresponding to the $NVT\DPsi$ ensemble. From Ref.~\citenum{scalfi_semiclassical_2020}, the extended TF Hamiltonian is
\begin{align}
\label{eq:hamiltonian_tf}
\mathH(\bfrN, \bfpN, \bfq) &= 
\mathK(\bfpN) + \mathU_0(\bfrN) + \frac{\bfqT\matAtf\bfq}{2} - \bfqT \bfB 
\; ,
\end{align}
where $\mathK$ is the kinetic energy and $\mathU_0$ contains the electrostatic interactions within the electrolyte, the non-electrostatic terms within the electrolyte and with the electrode atoms and a constant term involving the energy of the Fermi level. Furthermore, $\bfB$ is the vector of electrostatic potentials due to the electrolyte on each electrode atom and we introduced a modified symmetric matrix
\begin{equation}\label{eq:AdefTF}
\matAtf \equiv \mathbf{A}(\ltf) = \matA + \frac{\ltf^2 d}{\epsilon_0} \bfI \, ,
\end{equation}
with $\matA$ the symmetric $M \times M$ matrix describing the electrode-electrode electrostatic interactions for a perfect metal ($\ltf=0$), $d$ is the atomic density, $\epsilon_0$ the vacuum permittivity and $\bfI$ the $M\times M$ identity matrix.
Because the Hamiltonian is quadratic in the charges $\bfq$, the statistical mechanics framework derived in Ref.~\citenum{scalfi2020a} remains valid for the extended constant potential TF simulations. 

\newpage
\onecolumngrid

In the present work, we introduce a new thermodynamic integration approach in order to predict the evolution of the free energy as a function of the Thomas-Fermi length, $\ltf$, given by
\begin{equation}\label{eq:dftot}
\Delta F^{\DPsi}(\ltf) = F^{\DPsi}(\ltf) - F^{\DPsi}(0) = \Delta F^{\DPsi, BO}(\ltf) + \Delta F^{nBO}(\ltf) \, ,
\end{equation}
where the separation into Born-Oppenheimer (BO) and non-BO contributions is possible thanks to the factorization of the partition function $\mathZ = K \mathZ^{BO}$, where $K$ arises from the suppressed charged fluctuations in the BO approximation  (see Ref.~\citenum{scalfi2020a} for the derivation of this result in the case $\ltf=0$). Since the non-BO term depends only on the electrode, \emph{i.e.} neither on the presence or absence of electrolyte nor on voltage, it cancels in all differences considered in the main text.
We provide here a derivation of the expression given in the main text for the BO part of the free energy. We first write the free energy difference, at fixed voltage, between $\ltf=0$ (perfect metal) and a finite $\ltf$, as the integral of its derivative with respect to $\ltf$
\begin{equation}\label{eq:ti_ltf_def}
\Delta F^{\DPsi, BO}(\ltf) = F^{\DPsi, BO}(\ltf) - F^{\DPsi, BO}(0) = \int \limits _0 ^{\ltf} \mathrm{d}l \left( \frac{\partial F^{\DPsi,BO}}{\partial l}\right)_{NVT\DPsi} \, .
\end{equation}
Using the expression of the partition function in the BO ensemble (see Ref.~\citenum{scalfi2020a}), $\mathZ^{BO}$, the derivative of $F^{\DPsi,BO} = - \beta^{-1} \ln  \mathZ^{BO}$ with respect to $\ltf$ reads
\begin{align}\label{eq:derivative}
\left( \frac{\partial F^{\DPsi, BO}}{\partial \ltf}\right) &= - \beta ^{-1} (\mathZ^{BO})^{-1} \int\dbfrN\ e^{ -\beta \mathUz(\bfrN)} \frac{\beta}{2} \left( \frac{\partial \bfqsT\matAtf\bfqs}{\partial \ltf}\right) e^{\frac{\beta}{2} \bfqsT\matAtf\bfqs} 
&= -\frac{1}{2} \left< \left( \frac{\partial \bfqsT\matAtf\bfqs}{\partial \ltf}\right) \right> \, ,
\end{align}
where $\bfqs$ is the set of charges that enforce both the constant-potential and electroneutrality constraints
\begin{align}
\label{eq:qstar}
\bfqs(\bfrN) &= \matAinvtf \left[ \bfB +\bfPsi - \chi(\bfrN)\bfE \right]
\;,
\end{align}
where we used the Lagrange multiplier $\chi$ defined in Ref.~\citenum{scalfi2020a} and $\bfE= \{1, \dots, 1\}$ is a vector of size $M$.
Noticing that $\matAtf$ and $\bfqs$ depend on $\ltf$ explicitly but that $\bfB$ and $\bfPsi$ don't, and using standard rules for derivation and matrix algebra, we then write
\begin{align}
\frac{\partial}{\partial \ltf} \left[\bfqsT\matAtf\bfqs\right] &= \frac{\partial}{\partial \ltf} \left[\bfqsT\matAtf\matAinvtf\matAtf\bfqs\right]
 = \frac{\partial}{\partial \ltf} \left[ \matAtf\bfqs \right]^T \matAinvtf \left[ \matAtf\bfqs \right] \nonumber \\
 &= \left[ \frac{\partial (\matAtf\bfqs)}{\partial \ltf}  \right]^T \matAinvtf \left[ \matAtf\bfqs \right] +
 \left[ \matAtf\bfqs \right]^T \left(\frac{\partial \matAinvtf }{\partial \ltf} \right) \left[ \matAtf\bfqs \right] +  \left[ \matAtf\bfqs \right]^T \matAinvtf \left[ \frac{\partial (\matAtf\bfqs)}{\partial \ltf} \right] \nonumber \\
  &= \left[ -\frac{\partial \chi}{\partial \ltf} \bfE \right]^T \bfqs +
 \left[ \bfqsT\matAtf\right]\left(- \matAinvtf\frac{\partial \matAtf }{\partial \ltf}\matAinvtf \right) \left[ \matAtf\bfqs \right] +  \bfqsT\left[ -\frac{\partial \chi}{\partial \ltf} \bfE \right] \nonumber \\
 &= 0 +
 \bfqsT\left(- \frac{2\ltf d}{\epsilon_0}\bfI \right)\bfqs +  0
 = - \frac{2\ltf d}{\epsilon_0} \bfqsT \bfqs
\end{align}
where we used Eq.~\ref{eq:qstar} from the second to third line and the electroneutrality condition $\bfET\bfqs = 0$ from the third to fourth line. Introducing this result in Eqs.~\ref{eq:derivative} and~\ref{eq:ti_ltf_def}, we finally obtain:
\begin{equation}\label{eq:dfboti}
\Delta F^{\DPsi,BO}(\ltf) = F^{\DPsi,BO}(\ltf) - F^{\DPsi,BO}(0) = \int \limits _0 ^{\ltf} {\rm d} l \, \frac{l d}{\epsilon_0} \left< \bfqsT \bfqs \right>_{NVT\DPsi, l} \, .
\end{equation}
We note that this quantity is positive, \emph{i.e.} that the free energy increases from the perfect metal case to one charaterized by a finite $\ltf$.

\section{Thermodynamic cycle}
In the main text, we introduce various free energy differences associated with different processes, in particular changing the voltage $\DPsi$ or the Thomas-Fermi length $\ltf$. This is illustrated in the following thermodynamic cycle, changing the screening length vertically and charging or discharging the capacitor horizontally:
\begin{gather*}
\{\ltf=0, \DPsi=0\} 
\xrightarrow[]{\textstyle F_{SX}^{\DPsi}(\ltf=0) - F_{SX}^{0}(\ltf=0)} 
\{\ltf=0, \DPsi\} \\
\quad -\Delta F_{SX}^{0}(\ltf) \Bigg\uparrow 
\qquad \qquad \qquad \qquad \qquad \qquad \qquad
\Bigg\downarrow \Delta F_{SX}^{\DPsi}(\ltf) \\
\{\ltf, \DPsi=0\} \quad  
\xleftarrow[\textstyle -F_{SX}^{\DPsi}(\ltf) + F_{SX}^{0}(\ltf)]{} \quad 
\{\ltf, \DPsi\} 
\end{gather*}

\newpage
\twocolumngrid
where the subscripts $SX$ can refer to the empty capacitor ($SV=$~solid-vapor) or to the full electrochemical cell ($SL=$~solid-liquid). We also introduce the difference $\Delta\Delta F_{SX}^{\DPsi}(\ltf) = \Delta F_{SX}^{\DPsi}(\ltf) - \Delta F_{SV}^{0}(\ltf)$. For the empty capacitor with $\DPsi=0$~V, $\Delta F_{SV}^0 (\ltf)$ has no BO contribution because for zero voltage with no electrolyte, no charges are induced on the surface. It therefore follows that $\Delta F_{SV}^0 (\ltf) = \Delta F^{nBO}(\ltf)$ for the given set of (fixed) electrodes and, using Eq.~\ref{eq:dftot} above, that $\Delta\Delta F_{SX}^{\DPsi}(\ltf) = \Delta F_{SX}^{\DPsi, BO}(\ltf)$ corresponds to the sole BO contribution to the free energy difference, \emph{i.e.} that obtained directly from the BO sampling in constant-potential simulations (see Eq.~11 of the main text).

\section{Methods}

For constant-potential simulations, electrode atoms bear a Gaussian charge distribution of fixed width $\eta^{-1} = 0.56$~\AA\ and magnitude determined at each time step to enforce the constant potential and global electroneutrality constraints using a matrix inversion method (for electrochemical cells) or a conjugate gradient method (for empty capacitors and drop simulations). We consider TF lengths $\ltf$ ranging from 0.0 to 15.0~\AA\ for the empty capacitor and from 0.0 to 5.0~\AA\ in the presence of aqueous NaCl electrolyte, to ensure that the depth $na$ of the electrodes, with $n$ the number of atomic planes and $a$ the interplane distance, is larger than $\ltf$. We use two-dimensional boundary conditions (no periodicity in the $z$ direction), with 2D Ewald summation to compute electrostatic interactions in the presence of Gaussian charges~\cite{reed2007a,gingrich_ewald_2010} and a cutoff of 17.0~\AA\ for both the non-electrostatic interactions, described by truncated shifted Lennard-Jones (LJ) potentials, and the short range part of the Coulomb interactions. Water is modeled by the SPC/E force field~\cite{berendsen1987a}, while LJ parameters for Na$^+$ and Cl$^-$ are taken from Ref.~\citenum{dang_mechanism_1995} and thoses for carbon and gold electrode atoms are from Refs.~\citenum{werder_watercarbon_2003} and~\citenum{berg_evaluation_2017}, respectively; LJ parameters between different atom types are computed using the the Lorentz-Berthelot mixing rules.  

For graphite capacitors, each electrode consists of $n=50$ (resp. 5) planes for empty capacitors (resp. electrochemical cells) separated by $a=3.354$~\AA, with 480 carbon atoms per plane (surface area $\mathcal{A}=34.101\times36.915$~\AA$^2$); larger simulation boxes were also studied with $\mathcal{A}=68.202\times73.830$~\AA$^2$. Contact angle measurements are made by equilibrating a drop of electrolyte on a single electrode of surface area $\mathcal{A}=102.302\times110.745$~\AA$^2$ (4320 carbon atoms per plane) and only $n=3$ planes, using constant zero charges or a constant potential condition. The uncertainty reported on the contact angles is the standard error among 5 blocks of the simulations.
For gold-like electrodes, the box length in both the $x$ and $y$ directions is $L_{x} = L_y = 36.630$~\AA\ with 162 atoms per atomic plane. The structure is face-centered cubic with a lattice parameter of $4.07$~\AA\ and a (100) surface in contact with the electrolyte. The electrodes consist of $n= 10$ planes, separated by $a = 2.035$~\AA\, and held at a potential difference of $\DPsi = 0, 1$ or 2~V. In all cases, the electrolyte is composed of 2160 water molecules and 39 NaCl ion pairs. The capacitor simulation boxes are equilibrated at atmospheric pressure for 500~ps by applying a constant force to the electrodes (treated as rigid bodies) with $l_{TF}=0.0$~\AA; the electrodes separation is then fixed to the equilibrium value (for which the density in the middle of the liquid slab is equal to its bulk value) $L=56.2$~\AA\, for graphite electrodes and $L=50.7$~\AA\, for gold electrodes. For empty graphite capacitors we consider a range of distances from 20.0 to 300.0~\AA. Simulations are run with a timestep of 1~fs and the temperature is set at 298~K using a Nos\'e-Hoover chain thermostat. Capacitor simulations are run for at least 6~ns (those with the larger surfaces for 750~ps), while drop simulations are run for 3~ns. Differential capacitances are computed from the fluctuations of the total charge~\cite{limmer2013a,scalfi2020a} and errorbars estimated using the standard errors on the variance corrected for data correlations. All simulations are performed using the molecular dynamics code MetalWalls~\cite{marin-lafleche_metalwalls_2020}.

\section{Water structure in the first adsorbed layer on graphite}

Fig.~\ref{fig:gOO} reports the 2D oxygen-oxygen radial distribution function for water molecules in the first layer adsorbed on each electrode, for a range of $\ltf$, in a 1M NaCl / graphite capacitor. The results are identical for all the considered $\ltf$ values, with in particular a first maximum for a radial distance $\approx2.7$~\AA. Fig.~\ref{fig:worient} further shows that the orientation of water molecules in the same layers, quantified by the distribution $\cos\theta$, with $\theta$ the angle between the molecular dipole and the normal to the electrode surfaces also slightly depends on the Thomas-Fermi screening length.

\begin{figure}
\centering
\includegraphics[width=8.7cm]{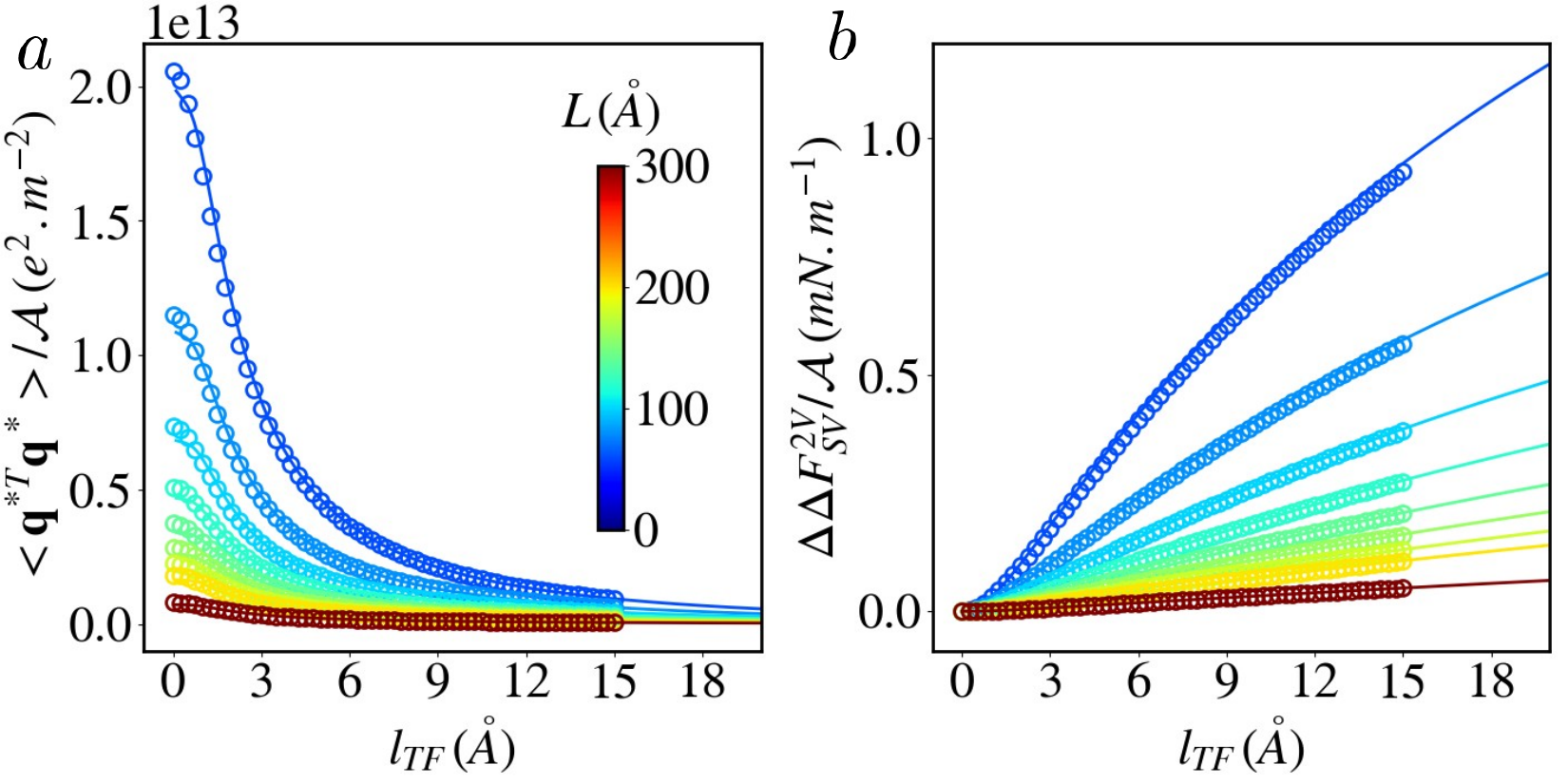}
\caption{(a) Average sum of the square of the atomic electrode charges, per unit area, as a function of $\ltf$. 
(b) Free energy difference per unit area $\Delta\Delta F_{SV}^{2V}(\ltf) / \mathcal{A}$ due to a change in the Thomas-Fermi length (see Eq.~6 of the main text), as a function of $\ltf$, computed from Eq.~\ref{eq:dfboti}.
Values are shown for an empty capacitor consisting of two graphite electrodes at $\DPsi = 2$~V, separated by a variable distance $L$ ranging from $60.0$ to $300.0$~\AA\, corresponding to different colors in panels a and b. Open circles are simulation data, while solid lines are the analytical expression Eq.~12 of the main text for panel b and the corresponding one for panel a.
\label{fig:tfti-2V}
}
\end{figure}

\begin{figure}
\centering
\includegraphics[width=8.7cm]{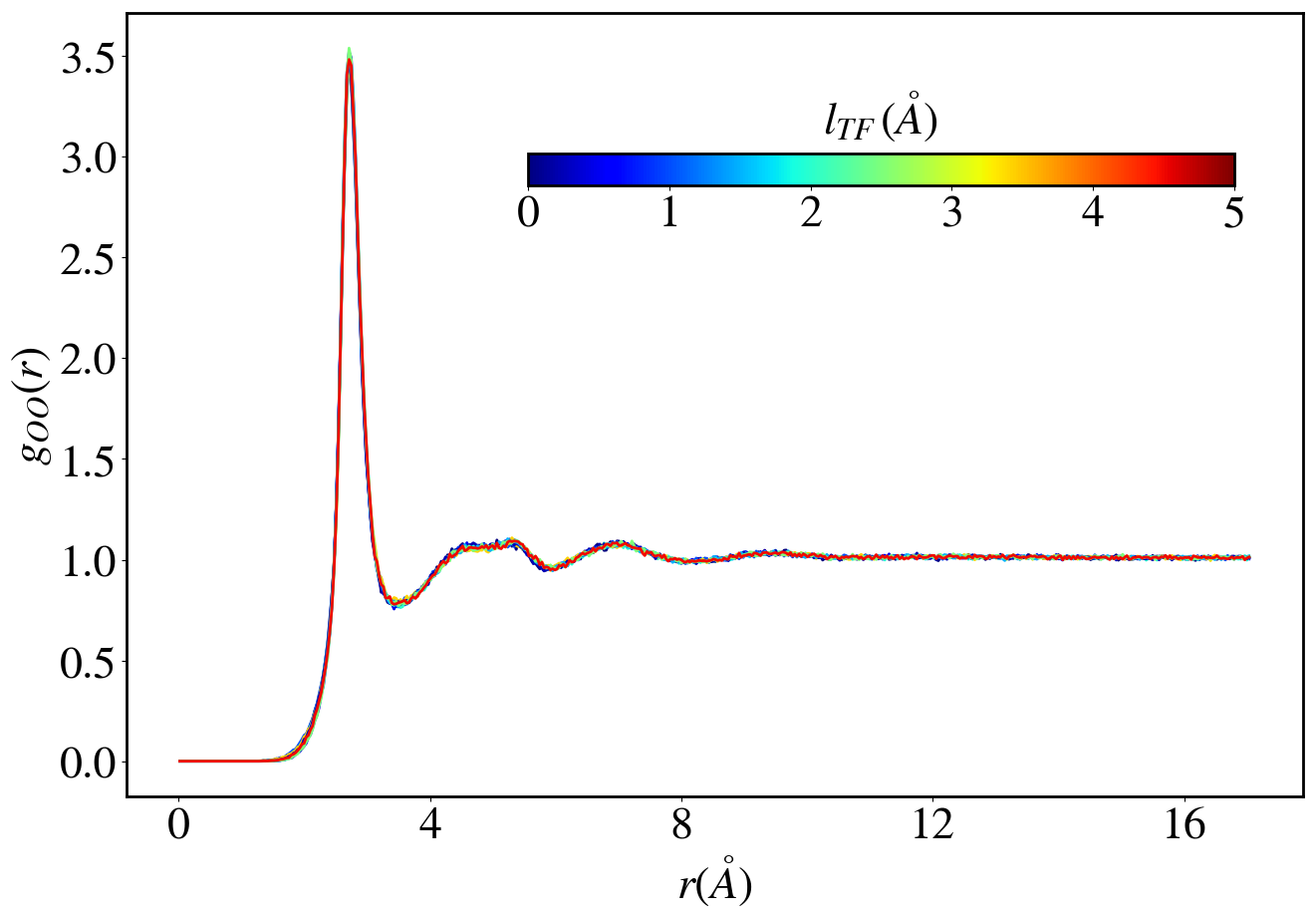}
\caption{Lateral correlations between water molecules in the first layer adsorbed on each electrode in a 1M NaCl / graphite capacitor at $\DPsi=0$~V. The figure shows the 2D radial distribution function of oxygen atoms, for a range of $\ltf$ indicated by the color bar. All curves are superimposed.
\label{fig:gOO}
}
\end{figure}

\begin{figure}
\centering
\includegraphics[width=8.7cm]{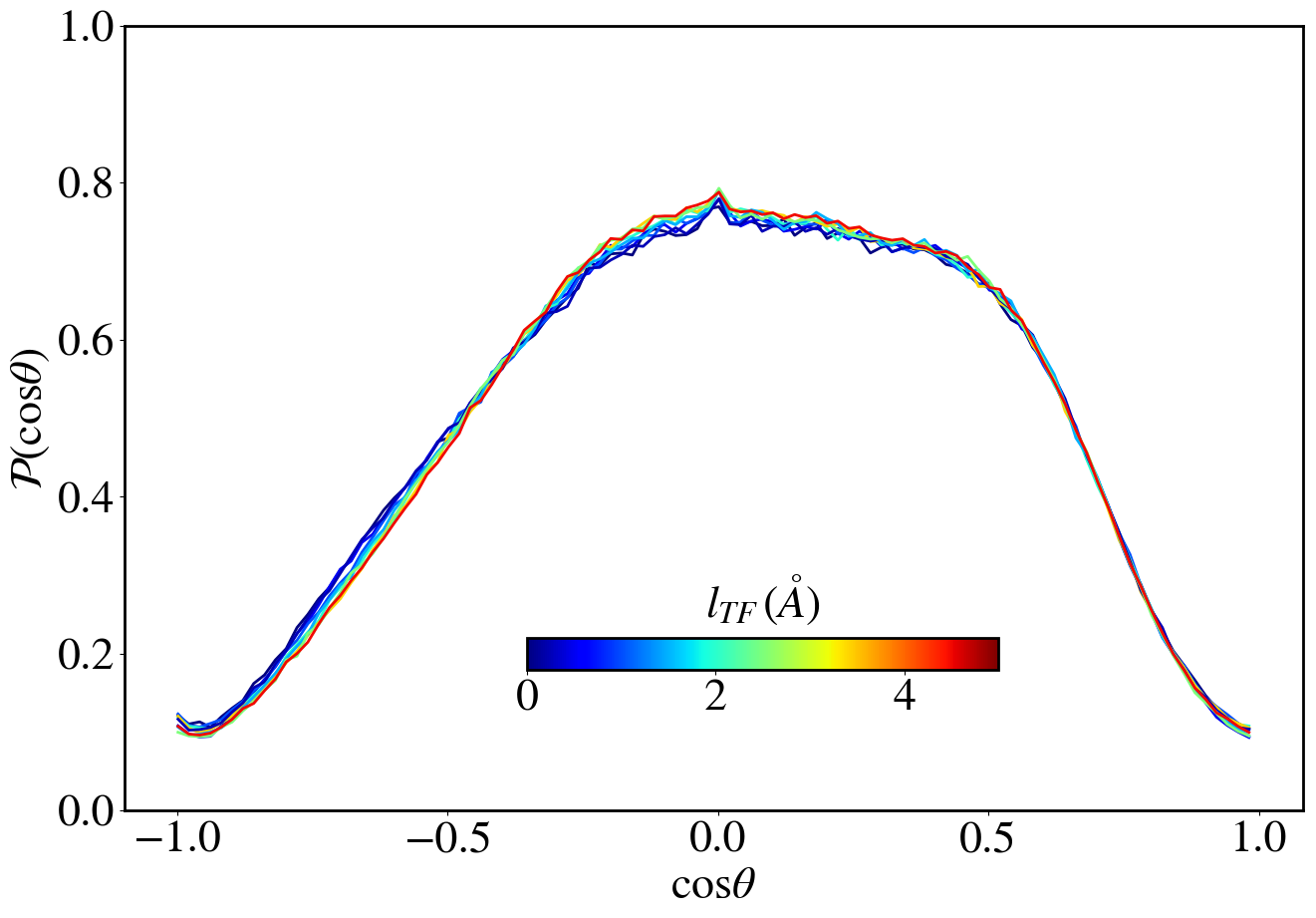}
\caption{Orientation of water molecules in the first layer adsorbed on each electrode in a 1M NaCl / graphite capacitor at $\DPsi=0$~V. The figure shows the distribution of $\cos\theta$, with $\theta$ the angle between the molecular dipole and the outward normal to the electrode surfaces, for a range of $\ltf$ indicated by the color bar.
\label{fig:worient}
}
\end{figure}


\begin{thebibliography}{48}%
\makeatletter
\providecommand \@ifxundefined [1]{%
 \@ifx{#1\undefined}
}%
\providecommand \@ifnum [1]{%
 \ifnum #1\expandafter \@firstoftwo
 \else \expandafter \@secondoftwo
 \fi
}%
\providecommand \@ifx [1]{%
 \ifx #1\expandafter \@firstoftwo
 \else \expandafter \@secondoftwo
 \fi
}%
\providecommand \natexlab [1]{#1}%
\providecommand \enquote  [1]{``#1''}%
\providecommand \bibnamefont  [1]{#1}%
\providecommand \bibfnamefont [1]{#1}%
\providecommand \citenamefont [1]{#1}%
\providecommand \href@noop [0]{\@secondoftwo}%
\providecommand \href [0]{\begingroup \@sanitize@url \@href}%
\providecommand \@href[1]{\@@startlink{#1}\@@href}%
\providecommand \@@href[1]{\endgroup#1\@@endlink}%
\providecommand \@sanitize@url [0]{\catcode `\\12\catcode `\$12\catcode
  `\&12\catcode `\#12\catcode `\^12\catcode `\_12\catcode `\%12\relax}%
\providecommand \@@startlink[1]{}%
\providecommand \@@endlink[0]{}%
\providecommand \url  [0]{\begingroup\@sanitize@url \@url }%
\providecommand \@url [1]{\endgroup\@href {#1}{\urlprefix }}%
\providecommand \urlprefix  [0]{URL }%
\providecommand \Eprint [0]{\href }%
\providecommand \doibase [0]{http://dx.doi.org/}%
\providecommand \selectlanguage [0]{\@gobble}%
\providecommand \bibinfo  [0]{\@secondoftwo}%
\providecommand \bibfield  [0]{\@secondoftwo}%
\providecommand \translation [1]{[#1]}%
\providecommand \BibitemOpen [0]{}%
\providecommand \bibitemStop [0]{}%
\providecommand \bibitemNoStop [0]{.\EOS\space}%
\providecommand \EOS [0]{\spacefactor3000\relax}%
\providecommand \BibitemShut  [1]{\csname bibitem#1\endcsname}%
\let\auto@bib@innerbib\@empty
\bibitem [{\citenamefont {Mugele}\ and\ \citenamefont
  {Baret}(2005)}]{mugele_electrowetting_2005}%
  \BibitemOpen
  \bibfield  {author} {\bibinfo {author} {\bibfnamefont {F.}~\bibnamefont
  {Mugele}}\ and\ \bibinfo {author} {\bibfnamefont {J.-C.}\ \bibnamefont
  {Baret}},\ }\bibfield  {title} {\enquote {\bibinfo {title} {Electrowetting:
  from basics to applications},}\ }\href {\doibase 10.1088/0953-8984/17/28/R01}
  {\bibfield  {journal} {\bibinfo  {journal} {Journal of Physics: Condensed
  Matter}\ }\textbf {\bibinfo {volume} {17}},\ \bibinfo {pages} {R705--R774}
  (\bibinfo {year} {2005})},\ \bibinfo {note} {publisher: IOP
  Publishing}\BibitemShut {NoStop}%
\bibitem [{\citenamefont {Daub}\ \emph {et~al.}(2007)\citenamefont {Daub},
  \citenamefont {Bratko}, \citenamefont {Leung},\ and\ \citenamefont
  {Luzar}}]{daub_electrowetting_2007}%
  \BibitemOpen
  \bibfield  {author} {\bibinfo {author} {\bibfnamefont {C.~D.}\ \bibnamefont
  {Daub}}, \bibinfo {author} {\bibfnamefont {D.}~\bibnamefont {Bratko}},
  \bibinfo {author} {\bibfnamefont {K.}~\bibnamefont {Leung}}, \ and\ \bibinfo
  {author} {\bibfnamefont {A.}~\bibnamefont {Luzar}},\ }\bibfield  {title}
  {{\selectlanguage {english}\enquote {\bibinfo {title} {Electrowetting at the
  {Nanoscale}},}\ }}\href {\doibase 10.1021/jp067395e} {\bibfield  {journal}
  {\bibinfo  {journal} {J. Phys. Chem. C}\ }\textbf {\bibinfo {volume} {111}},\
  \bibinfo {pages} {505--509} (\bibinfo {year} {2007})}\BibitemShut {NoStop}%
\bibitem [{\citenamefont {Choudhuri}\ \emph {et~al.}(2016)\citenamefont
  {Choudhuri}, \citenamefont {Vanzo}, \citenamefont {Madden}, \citenamefont
  {Salanne}, \citenamefont {Bratko},\ and\ \citenamefont
  {Luzar}}]{choudhuri_dynamic_2016}%
  \BibitemOpen
  \bibfield  {author} {\bibinfo {author} {\bibfnamefont {J.~R.}\ \bibnamefont
  {Choudhuri}}, \bibinfo {author} {\bibfnamefont {D.}~\bibnamefont {Vanzo}},
  \bibinfo {author} {\bibfnamefont {P.~A.}\ \bibnamefont {Madden}}, \bibinfo
  {author} {\bibfnamefont {M.}~\bibnamefont {Salanne}}, \bibinfo {author}
  {\bibfnamefont {D.}~\bibnamefont {Bratko}}, \ and\ \bibinfo {author}
  {\bibfnamefont {A.}~\bibnamefont {Luzar}},\ }\bibfield  {title}
  {{\selectlanguage {english}\enquote {\bibinfo {title} {Dynamic {Response} in
  {Nanoelectrowetting} on a {Dielectric}},}\ }}\href {\doibase
  10.1021/acsnano.6b03753} {\bibfield  {journal} {\bibinfo  {journal} {ACS
  Nano}\ }\textbf {\bibinfo {volume} {10}},\ \bibinfo {pages} {8536--8544}
  (\bibinfo {year} {2016})}\BibitemShut {NoStop}%
\bibitem [{\citenamefont {Sweeney}\ \emph {et~al.}(2012)\citenamefont
  {Sweeney}, \citenamefont {Hausen}, \citenamefont {Hayes}, \citenamefont
  {Webber}, \citenamefont {Endres}, \citenamefont {Rutland}, \citenamefont
  {Bennewitz},\ and\ \citenamefont {Atkin}}]{sweeney2012a}%
  \BibitemOpen
  \bibfield  {author} {\bibinfo {author} {\bibfnamefont {J.}~\bibnamefont
  {Sweeney}}, \bibinfo {author} {\bibfnamefont {F.}~\bibnamefont {Hausen}},
  \bibinfo {author} {\bibfnamefont {R.}~\bibnamefont {Hayes}}, \bibinfo
  {author} {\bibfnamefont {G.~B.}\ \bibnamefont {Webber}}, \bibinfo {author}
  {\bibfnamefont {F.}~\bibnamefont {Endres}}, \bibinfo {author} {\bibfnamefont
  {M.~W.}\ \bibnamefont {Rutland}}, \bibinfo {author} {\bibfnamefont
  {R.}~\bibnamefont {Bennewitz}}, \ and\ \bibinfo {author} {\bibfnamefont
  {R.}~\bibnamefont {Atkin}},\ }\bibfield  {title} {\enquote {\bibinfo {title}
  {Control of nanoscale friction on gold in an ionic liquid by a
  potential-dependent ionic lubricant layer},}\ }\href@noop {} {\bibfield
  {journal} {\bibinfo  {journal} {Phys. Rev. Lett.}\ }\textbf {\bibinfo
  {volume} {109}},\ \bibinfo {pages} {155502} (\bibinfo {year}
  {2012})}\BibitemShut {NoStop}%
\bibitem [{\citenamefont {Li}\ \emph {et~al.}(2014)\citenamefont {Li},
  \citenamefont {Wood}, \citenamefont {Rutland},\ and\ \citenamefont
  {Atkin}}]{li_ionic_2014}%
  \BibitemOpen
  \bibfield  {author} {\bibinfo {author} {\bibfnamefont {H.}~\bibnamefont
  {Li}}, \bibinfo {author} {\bibfnamefont {R.~J.}\ \bibnamefont {Wood}},
  \bibinfo {author} {\bibfnamefont {M.~W.}\ \bibnamefont {Rutland}}, \ and\
  \bibinfo {author} {\bibfnamefont {R.}~\bibnamefont {Atkin}},\ }\bibfield
  {title} {\enquote {\bibinfo {title} {An ionic liquid lubricant enables
  superlubricity to be “switched on” in situ using an electrical
  potential},}\ }\href {\doibase 10.1039/C4CC00979G} {\bibfield  {journal}
  {\bibinfo  {journal} {Chemical Communications}\ }\textbf {\bibinfo {volume}
  {50}},\ \bibinfo {pages} {4368--4370} (\bibinfo {year} {2014})}\BibitemShut
  {NoStop}%
\bibitem [{\citenamefont {Fajardo}\ \emph
  {et~al.}(2015{\natexlab{a}})\citenamefont {Fajardo}, \citenamefont {Bresme},
  \citenamefont {Kornyshev},\ and\ \citenamefont {Urbakh}}]{fajardo2015a}%
  \BibitemOpen
  \bibfield  {author} {\bibinfo {author} {\bibfnamefont {O.~Y.}\ \bibnamefont
  {Fajardo}}, \bibinfo {author} {\bibfnamefont {F.}~\bibnamefont {Bresme}},
  \bibinfo {author} {\bibfnamefont {A.~A.}\ \bibnamefont {Kornyshev}}, \ and\
  \bibinfo {author} {\bibfnamefont {M.}~\bibnamefont {Urbakh}},\ }\bibfield
  {title} {\enquote {\bibinfo {title} {Electrotunable lubricity with ionic
  liquid nanoscale films},}\ }\href@noop {} {\bibfield  {journal} {\bibinfo
  {journal} {Sci. Rep.}\ }\textbf {\bibinfo {volume} {5}},\ \bibinfo {pages}
  {7698} (\bibinfo {year} {2015}{\natexlab{a}})}\BibitemShut {NoStop}%
\bibitem [{\citenamefont {Fajardo}\ \emph
  {et~al.}(2015{\natexlab{b}})\citenamefont {Fajardo}, \citenamefont {Bresme},
  \citenamefont {Kornyshev},\ and\ \citenamefont {Urbakh}}]{fajardo2015b}%
  \BibitemOpen
  \bibfield  {author} {\bibinfo {author} {\bibfnamefont {O.~Y.}\ \bibnamefont
  {Fajardo}}, \bibinfo {author} {\bibfnamefont {F.}~\bibnamefont {Bresme}},
  \bibinfo {author} {\bibfnamefont {A.~A.}\ \bibnamefont {Kornyshev}}, \ and\
  \bibinfo {author} {\bibfnamefont {M.}~\bibnamefont {Urbakh}},\ }\bibfield
  {title} {\enquote {\bibinfo {title} {Electrotunable friction with ionic
  liquid lubricants: How important is the molecular structure of the ions?}}\
  }\href@noop {} {\bibfield  {journal} {\bibinfo  {journal} {J. Phys. Chem.
  Lett.}\ }\textbf {\bibinfo {volume} {6}},\ \bibinfo {pages} {3998} (\bibinfo
  {year} {2015}{\natexlab{b}})}\BibitemShut {NoStop}%
\bibitem [{\citenamefont {Pivnic}\ \emph {et~al.}(2020)\citenamefont {Pivnic},
  \citenamefont {Bresme}, \citenamefont {Kornyshev},\ and\ \citenamefont
  {Urbakh}}]{pivnic_electrotunable_2020}%
  \BibitemOpen
  \bibfield  {author} {\bibinfo {author} {\bibfnamefont {K.}~\bibnamefont
  {Pivnic}}, \bibinfo {author} {\bibfnamefont {F.}~\bibnamefont {Bresme}},
  \bibinfo {author} {\bibfnamefont {A.~A.}\ \bibnamefont {Kornyshev}}, \ and\
  \bibinfo {author} {\bibfnamefont {M.}~\bibnamefont {Urbakh}},\ }\bibfield
  {title} {\enquote {\bibinfo {title} {Electrotunable {Friction} in {Diluted}
  {Room} {Temperature} {Ionic} {Liquids}: {Implications} for
  {Nanotribology}},}\ }\href {\doibase 10.1021/acsanm.0c01946} {\bibfield
  {journal} {\bibinfo  {journal} {ACS Applied Nano Materials}\ } (\bibinfo
  {year} {2020}),\ 10.1021/acsanm.0c01946}\BibitemShut {NoStop}%
\bibitem [{\citenamefont {Perez-Martinez}\ and\ \citenamefont
  {Perkin}(2019)}]{perez-martinez_surface_2019}%
  \BibitemOpen
  \bibfield  {author} {\bibinfo {author} {\bibfnamefont {C.~S.}\ \bibnamefont
  {Perez-Martinez}}\ and\ \bibinfo {author} {\bibfnamefont {S.}~\bibnamefont
  {Perkin}},\ }\bibfield  {title} {\enquote {\bibinfo {title} {Surface forces
  generated by the action of electric fields across liquid films},}\ }\href
  {\doibase 10.1039/C9SM00143C} {\bibfield  {journal} {\bibinfo  {journal}
  {Soft Matter}\ }\textbf {\bibinfo {volume} {15}},\ \bibinfo {pages}
  {4255--4265} (\bibinfo {year} {2019})}\BibitemShut {NoStop}%
\bibitem [{\citenamefont {Comtet}\ \emph {et~al.}(2017)\citenamefont {Comtet},
  \citenamefont {Nigu\`es}, \citenamefont {Kaiser}, \citenamefont {Coasne},
  \citenamefont {Bocquet},\ and\ \citenamefont {Siria}}]{comtet2017a}%
  \BibitemOpen
  \bibfield  {author} {\bibinfo {author} {\bibfnamefont {J.}~\bibnamefont
  {Comtet}}, \bibinfo {author} {\bibfnamefont {A.}~\bibnamefont {Nigu\`es}},
  \bibinfo {author} {\bibfnamefont {V.}~\bibnamefont {Kaiser}}, \bibinfo
  {author} {\bibfnamefont {B.}~\bibnamefont {Coasne}}, \bibinfo {author}
  {\bibfnamefont {L.}~\bibnamefont {Bocquet}}, \ and\ \bibinfo {author}
  {\bibfnamefont {A.}~\bibnamefont {Siria}},\ }\bibfield  {title} {\enquote
  {\bibinfo {title} {Nanoscale capillary freezing of ionic liquids confined
  between metallic interfaces and the role of electronic screening},}\
  }\href@noop {} {\bibfield  {journal} {\bibinfo  {journal} {Nat. Mater.}\
  }\textbf {\bibinfo {volume} {16}},\ \bibinfo {pages} {634--639} (\bibinfo
  {year} {2017})}\BibitemShut {NoStop}%
\bibitem [{\citenamefont {Lainé}\ \emph {et~al.}(2020)\citenamefont {Lainé},
  \citenamefont {Niguès}, \citenamefont {Bocquet},\ and\ \citenamefont
  {Siria}}]{laine_nanotribology_2020}%
  \BibitemOpen
  \bibfield  {author} {\bibinfo {author} {\bibfnamefont {A.}~\bibnamefont
  {Lainé}}, \bibinfo {author} {\bibfnamefont {A.}~\bibnamefont {Niguès}},
  \bibinfo {author} {\bibfnamefont {L.}~\bibnamefont {Bocquet}}, \ and\
  \bibinfo {author} {\bibfnamefont {A.}~\bibnamefont {Siria}},\ }\bibfield
  {title} {\enquote {\bibinfo {title} {Nanotribology of {Ionic} {Liquids}:
  {Transition} to {Yielding} {Response} in {Nanometric} {Confinement} with
  {Metallic} {Surfaces}},}\ }\href {\doibase 10.1103/PhysRevX.10.011068}
  {\bibfield  {journal} {\bibinfo  {journal} {Physical Review X}\ }\textbf
  {\bibinfo {volume} {10}},\ \bibinfo {pages} {011068} (\bibinfo {year}
  {2020})},\ \bibinfo {note} {publisher: American Physical Society}\BibitemShut
  {NoStop}%
\bibitem [{\citenamefont {Garcia}\ \emph {et~al.}(2017)\citenamefont {Garcia},
  \citenamefont {Jacquot}, \citenamefont {Charlaix},\ and\ \citenamefont
  {Cross}}]{garcia_nano-mechanics_2017}%
  \BibitemOpen
  \bibfield  {author} {\bibinfo {author} {\bibfnamefont {L.}~\bibnamefont
  {Garcia}}, \bibinfo {author} {\bibfnamefont {L.}~\bibnamefont {Jacquot}},
  \bibinfo {author} {\bibfnamefont {E.}~\bibnamefont {Charlaix}}, \ and\
  \bibinfo {author} {\bibfnamefont {B.}~\bibnamefont {Cross}},\ }\bibfield
  {title} {\enquote {\bibinfo {title} {Nano-mechanics of ionic liquids at
  dielectric and metallic interfaces},}\ }\href {\doibase 10.1039/C7FD00149E}
  {\bibfield  {journal} {\bibinfo  {journal} {Faraday Discussions}\ }\textbf
  {\bibinfo {volume} {206}},\ \bibinfo {pages} {443--457} (\bibinfo {year}
  {2017})}\BibitemShut {NoStop}%
\bibitem [{\citenamefont {Netz}(1999)}]{netz_debye-huckel_1999}%
  \BibitemOpen
  \bibfield  {author} {\bibinfo {author} {\bibfnamefont {R.~R.}\ \bibnamefont
  {Netz}},\ }\bibfield  {title} {\enquote {\bibinfo {title} {Debye-{H}\"uckel
  theory for interfacial geometries},}\ }\href {\doibase
  10.1103/PhysRevE.60.3174} {\bibfield  {journal} {\bibinfo  {journal}
  {Physical Review E}\ }\textbf {\bibinfo {volume} {60}},\ \bibinfo {pages}
  {3174--3182} (\bibinfo {year} {1999})}\BibitemShut {NoStop}%
\bibitem [{\citenamefont {Arnold}\ \emph {et~al.}(2013)\citenamefont {Arnold},
  \citenamefont {Breitsprecher}, \citenamefont {Fahrenberger}, \citenamefont
  {Kesselheim}, \citenamefont {Lenz},\ and\ \citenamefont
  {Holm}}]{arnold2013a}%
  \BibitemOpen
  \bibfield  {author} {\bibinfo {author} {\bibfnamefont {A.}~\bibnamefont
  {Arnold}}, \bibinfo {author} {\bibfnamefont {K.}~\bibnamefont
  {Breitsprecher}}, \bibinfo {author} {\bibfnamefont {F.}~\bibnamefont
  {Fahrenberger}}, \bibinfo {author} {\bibfnamefont {S.}~\bibnamefont
  {Kesselheim}}, \bibinfo {author} {\bibfnamefont {O.}~\bibnamefont {Lenz}}, \
  and\ \bibinfo {author} {\bibfnamefont {C.}~\bibnamefont {Holm}},\ }\bibfield
  {title} {\enquote {\bibinfo {title} {Efficient algorithms for electrostatic
  interactions including dielectric contrasts},}\ }\href@noop {} {\bibfield
  {journal} {\bibinfo  {journal} {Entropy}\ }\textbf {\bibinfo {volume} {15}},\
  \bibinfo {pages} {4569--4588} (\bibinfo {year} {2013})}\BibitemShut {NoStop}%
\bibitem [{\citenamefont {Breitsprecher}, \citenamefont {Szuttor},\ and\
  \citenamefont {Holm}(2015)}]{breitsprecher_electrode_2015}%
  \BibitemOpen
  \bibfield  {author} {\bibinfo {author} {\bibfnamefont {K.}~\bibnamefont
  {Breitsprecher}}, \bibinfo {author} {\bibfnamefont {K.}~\bibnamefont
  {Szuttor}}, \ and\ \bibinfo {author} {\bibfnamefont {C.}~\bibnamefont
  {Holm}},\ }\bibfield  {title} {{\selectlanguage {english}\enquote {\bibinfo
  {title} {Electrode {Models} for {Ionic} {Liquid}-{Based} {Capacitors}},}\
  }}\href {\doibase 10.1021/acs.jpcc.5b06046} {\bibfield  {journal} {\bibinfo
  {journal} {J. Phys. Chem. C}\ }\textbf {\bibinfo {volume} {119}},\ \bibinfo
  {pages} {22445--22451} (\bibinfo {year} {2015})}\BibitemShut {NoStop}%
\bibitem [{\citenamefont {Kornyshev}, \citenamefont {Luque},\ and\
  \citenamefont {Schmickler}(2014)}]{kornyshev2013a}%
  \BibitemOpen
  \bibfield  {author} {\bibinfo {author} {\bibfnamefont {A.~A.}\ \bibnamefont
  {Kornyshev}}, \bibinfo {author} {\bibfnamefont {N.~B.}\ \bibnamefont
  {Luque}}, \ and\ \bibinfo {author} {\bibfnamefont {W.}~\bibnamefont
  {Schmickler}},\ }\bibfield  {title} {\enquote {\bibinfo {title} {Differential
  capacitance of ionic liquid interface with graphite: the story of two double
  layers},}\ }\href@noop {} {\bibfield  {journal} {\bibinfo  {journal} {J.
  Solid State Electrochem.}\ }\textbf {\bibinfo {volume} {18}},\ \bibinfo
  {pages} {1345--1349} (\bibinfo {year} {2014})}\BibitemShut {NoStop}%
\bibitem [{\citenamefont {Lee}\ and\ \citenamefont {Perkin}(2016)}]{lee2016b}%
  \BibitemOpen
  \bibfield  {author} {\bibinfo {author} {\bibfnamefont {A.~A.}\ \bibnamefont
  {Lee}}\ and\ \bibinfo {author} {\bibfnamefont {S.}~\bibnamefont {Perkin}},\
  }\bibfield  {title} {\enquote {\bibinfo {title} {Ion-image interactions and
  phase transition at electrolyte-metal interfaces},}\ }\href@noop {}
  {\bibfield  {journal} {\bibinfo  {journal} {J. Phys. Chem. Lett.}\ }\textbf
  {\bibinfo {volume} {7}},\ \bibinfo {pages} {2753} (\bibinfo {year}
  {2016})}\BibitemShut {NoStop}%
\bibitem [{\citenamefont {Kornyshev}\ and\ \citenamefont
  {Vorotyntsev}(1980)}]{kornyshev_nonlocal_1980}%
  \BibitemOpen
  \bibfield  {author} {\bibinfo {author} {\bibfnamefont {A.}~\bibnamefont
  {Kornyshev}}\ and\ \bibinfo {author} {\bibfnamefont {M.}~\bibnamefont
  {Vorotyntsev}},\ }\bibfield  {title} {{\selectlanguage {english}\enquote
  {\bibinfo {title} {Nonlocal electrostatic approach to the double layer and
  adsorption at the electrode-electrolyte interface},}\ }}\href {\doibase
  10.1016/0039-6028(80)90597-X} {\bibfield  {journal} {\bibinfo  {journal}
  {Surface Science}\ }\textbf {\bibinfo {volume} {101}},\ \bibinfo {pages}
  {23--48} (\bibinfo {year} {1980})}\BibitemShut {NoStop}%
\bibitem [{\citenamefont {Kornyshev}, \citenamefont {Schmickler},\ and\
  \citenamefont {Vorotyntsev}(1982)}]{kornyshev_nonlocal_1982}%
  \BibitemOpen
  \bibfield  {author} {\bibinfo {author} {\bibfnamefont {A.~A.}\ \bibnamefont
  {Kornyshev}}, \bibinfo {author} {\bibfnamefont {W.}~\bibnamefont
  {Schmickler}}, \ and\ \bibinfo {author} {\bibfnamefont {M.~A.}\ \bibnamefont
  {Vorotyntsev}},\ }\bibfield  {title} {{\selectlanguage {english}\enquote
  {\bibinfo {title} {Nonlocal electrostatic approach to the problem of a double
  layer at a metal-electrolyte interface},}\ }}\href {\doibase
  10.1103/PhysRevB.25.5244} {\bibfield  {journal} {\bibinfo  {journal} {Phys.
  Rev. B}\ }\textbf {\bibinfo {volume} {25}},\ \bibinfo {pages} {5244--5256}
  (\bibinfo {year} {1982})}\BibitemShut {NoStop}%
\bibitem [{\citenamefont {Rochester}\ \emph {et~al.}(2013)\citenamefont
  {Rochester}, \citenamefont {Lee}, \citenamefont {Pruessner},\ and\
  \citenamefont {Kornyshev}}]{rochester2013a}%
  \BibitemOpen
  \bibfield  {author} {\bibinfo {author} {\bibfnamefont {C.~C.}\ \bibnamefont
  {Rochester}}, \bibinfo {author} {\bibfnamefont {A.~A.}\ \bibnamefont {Lee}},
  \bibinfo {author} {\bibfnamefont {G.}~\bibnamefont {Pruessner}}, \ and\
  \bibinfo {author} {\bibfnamefont {A.~A.}\ \bibnamefont {Kornyshev}},\
  }\bibfield  {title} {\enquote {\bibinfo {title} {Interionic interactions in
  conducting nanoconfinement},}\ }\href@noop {} {\bibfield  {journal} {\bibinfo
   {journal} {ChemPhysChem}\ }\textbf {\bibinfo {volume} {14}},\ \bibinfo
  {pages} {4121--4125} (\bibinfo {year} {2013})}\BibitemShut {NoStop}%
\bibitem [{\citenamefont {Kaiser}\ \emph {et~al.}(2017)\citenamefont {Kaiser},
  \citenamefont {Comtet}, \citenamefont {Nigu\`es}, \citenamefont {Siria},
  \citenamefont {Coasne},\ and\ \citenamefont {Bocquet}}]{kaiser2017a}%
  \BibitemOpen
  \bibfield  {author} {\bibinfo {author} {\bibfnamefont {V.}~\bibnamefont
  {Kaiser}}, \bibinfo {author} {\bibfnamefont {J.}~\bibnamefont {Comtet}},
  \bibinfo {author} {\bibfnamefont {A.}~\bibnamefont {Nigu\`es}}, \bibinfo
  {author} {\bibfnamefont {A.}~\bibnamefont {Siria}}, \bibinfo {author}
  {\bibfnamefont {B.}~\bibnamefont {Coasne}}, \ and\ \bibinfo {author}
  {\bibfnamefont {L.}~\bibnamefont {Bocquet}},\ }\bibfield  {title} {\enquote
  {\bibinfo {title} {Electrostatic interactions between ions near
  {Thomas-Fermi} substrates and the surface energy of ionic crystal at
  imperfect metals},}\ }\href@noop {} {\bibfield  {journal} {\bibinfo
  {journal} {Faraday Discuss.}\ }\textbf {\bibinfo {volume} {199}},\ \bibinfo
  {pages} {129--158} (\bibinfo {year} {2017})}\BibitemShut {NoStop}%
\bibitem [{\citenamefont {Siepmann}\ and\ \citenamefont
  {Sprik}(1995)}]{siepmann1995a}%
  \BibitemOpen
  \bibfield  {author} {\bibinfo {author} {\bibfnamefont {J.~I.}\ \bibnamefont
  {Siepmann}}\ and\ \bibinfo {author} {\bibfnamefont {M.}~\bibnamefont
  {Sprik}},\ }\bibfield  {title} {\enquote {\bibinfo {title} {{Influence of
  Surface-Topology and Electrostatic Potential on Water Electrode Systems}},}\
  }\href@noop {} {\bibfield  {journal} {\bibinfo  {journal} {J. Chem. Phys.}\
  }\textbf {\bibinfo {volume} {102}},\ \bibinfo {pages} {511--524} (\bibinfo
  {year} {1995})}\BibitemShut {NoStop}%
\bibitem [{\citenamefont {Reed}, \citenamefont {Lanning},\ and\ \citenamefont
  {Madden}(2007)}]{reed2007a}%
  \BibitemOpen
  \bibfield  {author} {\bibinfo {author} {\bibfnamefont {S.~K.}\ \bibnamefont
  {Reed}}, \bibinfo {author} {\bibfnamefont {O.~J.}\ \bibnamefont {Lanning}}, \
  and\ \bibinfo {author} {\bibfnamefont {P.~A.}\ \bibnamefont {Madden}},\
  }\bibfield  {title} {\enquote {\bibinfo {title} {{Electrochemical Interface
  Between an Ionic Liquid and a Model Metallic Electrode}},}\ }\href@noop {}
  {\bibfield  {journal} {\bibinfo  {journal} {J. Chem. Phys.}\ }\textbf
  {\bibinfo {volume} {126}},\ \bibinfo {pages} {084704} (\bibinfo {year}
  {2007})}\BibitemShut {NoStop}%
\bibitem [{\citenamefont {Limmer}\ \emph
  {et~al.}(2013{\natexlab{a}})\citenamefont {Limmer}, \citenamefont {Merlet},
  \citenamefont {Salanne}, \citenamefont {Chandler}, \citenamefont {Madden},
  \citenamefont {{van Roij}},\ and\ \citenamefont {Rotenberg}}]{limmer2013a}%
  \BibitemOpen
  \bibfield  {author} {\bibinfo {author} {\bibfnamefont {D.~T.}\ \bibnamefont
  {Limmer}}, \bibinfo {author} {\bibfnamefont {C.}~\bibnamefont {Merlet}},
  \bibinfo {author} {\bibfnamefont {M.}~\bibnamefont {Salanne}}, \bibinfo
  {author} {\bibfnamefont {D.}~\bibnamefont {Chandler}}, \bibinfo {author}
  {\bibfnamefont {P.~A.}\ \bibnamefont {Madden}}, \bibinfo {author}
  {\bibfnamefont {R.}~\bibnamefont {{van Roij}}}, \ and\ \bibinfo {author}
  {\bibfnamefont {B.}~\bibnamefont {Rotenberg}},\ }\bibfield  {title} {\enquote
  {\bibinfo {title} {Charge fluctuations in nanoscale capacitors},}\
  }\href@noop {} {\bibfield  {journal} {\bibinfo  {journal} {Phys. Rev. Lett.}\
  }\textbf {\bibinfo {volume} {111}},\ \bibinfo {pages} {106102} (\bibinfo
  {year} {2013}{\natexlab{a}})}\BibitemShut {NoStop}%
\bibitem [{\citenamefont {Scalfi}\ \emph
  {et~al.}(2020{\natexlab{a}})\citenamefont {Scalfi}, \citenamefont {Limmer},
  \citenamefont {Coretti}, \citenamefont {Bonella}, \citenamefont {Madden},
  \citenamefont {Salanne},\ and\ \citenamefont {Rotenberg}}]{scalfi2020a}%
  \BibitemOpen
  \bibfield  {author} {\bibinfo {author} {\bibfnamefont {L.}~\bibnamefont
  {Scalfi}}, \bibinfo {author} {\bibfnamefont {D.~T.}\ \bibnamefont {Limmer}},
  \bibinfo {author} {\bibfnamefont {A.}~\bibnamefont {Coretti}}, \bibinfo
  {author} {\bibfnamefont {S.}~\bibnamefont {Bonella}}, \bibinfo {author}
  {\bibfnamefont {P.~A.}\ \bibnamefont {Madden}}, \bibinfo {author}
  {\bibfnamefont {M.}~\bibnamefont {Salanne}}, \ and\ \bibinfo {author}
  {\bibfnamefont {B.}~\bibnamefont {Rotenberg}},\ }\bibfield  {title} {\enquote
  {\bibinfo {title} {Charge fluctuations from molecular simulations in the
  constant-potential ensemble},}\ }\href@noop {} {\bibfield  {journal}
  {\bibinfo  {journal} {Phys. Chem. Chem. Phys.}\ }\textbf {\bibinfo {volume}
  {22}},\ \bibinfo {pages} {10480--10489} (\bibinfo {year}
  {2020}{\natexlab{a}})}\BibitemShut {NoStop}%
\bibitem [{\citenamefont {Willard}\ \emph {et~al.}(2009)\citenamefont
  {Willard}, \citenamefont {Reed}, \citenamefont {Madden},\ and\ \citenamefont
  {Chandler}}]{willard2009a}%
  \BibitemOpen
  \bibfield  {author} {\bibinfo {author} {\bibfnamefont {A.~P.}\ \bibnamefont
  {Willard}}, \bibinfo {author} {\bibfnamefont {S.~K.}\ \bibnamefont {Reed}},
  \bibinfo {author} {\bibfnamefont {P.~A.}\ \bibnamefont {Madden}}, \ and\
  \bibinfo {author} {\bibfnamefont {D.}~\bibnamefont {Chandler}},\ }\bibfield
  {title} {\enquote {\bibinfo {title} {Water at an electrochemical interface -
  a simulation study},}\ }\href@noop {} {\bibfield  {journal} {\bibinfo
  {journal} {Faraday Discuss.}\ }\textbf {\bibinfo {volume} {141}},\ \bibinfo
  {pages} {423--441} (\bibinfo {year} {2009})}\BibitemShut {NoStop}%
\bibitem [{\citenamefont {Merlet}\ \emph
  {et~al.}(2013{\natexlab{a}})\citenamefont {Merlet}, \citenamefont
  {Rotenberg}, \citenamefont {Madden},\ and\ \citenamefont
  {Salanne}}]{merlet2013c}%
  \BibitemOpen
  \bibfield  {author} {\bibinfo {author} {\bibfnamefont {C.}~\bibnamefont
  {Merlet}}, \bibinfo {author} {\bibfnamefont {B.}~\bibnamefont {Rotenberg}},
  \bibinfo {author} {\bibfnamefont {P.~A.}\ \bibnamefont {Madden}}, \ and\
  \bibinfo {author} {\bibfnamefont {M.}~\bibnamefont {Salanne}},\ }\bibfield
  {title} {\enquote {\bibinfo {title} {Computer simulations of ionic liquids at
  electrochemical interfaces},}\ }\href@noop {} {\bibfield  {journal} {\bibinfo
   {journal} {Phys. Chem. Chem. Phys.}\ }\textbf {\bibinfo {volume} {15}},\
  \bibinfo {pages} {15781--15792} (\bibinfo {year}
  {2013}{\natexlab{a}})}\BibitemShut {NoStop}%
\bibitem [{\citenamefont {Pounds}, \citenamefont {Salanne},\ and\ \citenamefont
  {Madden}(2015)}]{pounds2015a}%
  \BibitemOpen
  \bibfield  {author} {\bibinfo {author} {\bibfnamefont {M.~A.}\ \bibnamefont
  {Pounds}}, \bibinfo {author} {\bibfnamefont {M.}~\bibnamefont {Salanne}}, \
  and\ \bibinfo {author} {\bibfnamefont {P.~A.}\ \bibnamefont {Madden}},\
  }\bibfield  {title} {\enquote {\bibinfo {title} {Molecular aspects of the
  {Eu}$^{3+}$/{Eu}$^{2+}$ redox reaction at the interface between a molten salt
  and a metallic electrode},}\ }\href@noop {} {\bibfield  {journal} {\bibinfo
  {journal} {Mol. Phys.}\ }\textbf {\bibinfo {volume} {113}},\ \bibinfo {pages}
  {2451--2462} (\bibinfo {year} {2015})}\BibitemShut {NoStop}%
\bibitem [{\citenamefont {Limmer}\ \emph
  {et~al.}(2013{\natexlab{b}})\citenamefont {Limmer}, \citenamefont {Willard},
  \citenamefont {Madden},\ and\ \citenamefont {Chandler}}]{limmer2013b}%
  \BibitemOpen
  \bibfield  {author} {\bibinfo {author} {\bibfnamefont {D.~T.}\ \bibnamefont
  {Limmer}}, \bibinfo {author} {\bibfnamefont {A.~P.}\ \bibnamefont {Willard}},
  \bibinfo {author} {\bibfnamefont {P.}~\bibnamefont {Madden}}, \ and\ \bibinfo
  {author} {\bibfnamefont {D.}~\bibnamefont {Chandler}},\ }\bibfield  {title}
  {\enquote {\bibinfo {title} {Hydration of metal surfaces can be dynamically
  heterogeneous and hydrophobic},}\ }\href@noop {} {\bibfield  {journal}
  {\bibinfo  {journal} {Proc. Natl. Acad. Sci. U.S.A.}\ }\textbf {\bibinfo
  {volume} {110}},\ \bibinfo {pages} {4200--4205} (\bibinfo {year}
  {2013}{\natexlab{b}})}\BibitemShut {NoStop}%
\bibitem [{\citenamefont {Willard}\ \emph {et~al.}(2013)\citenamefont
  {Willard}, \citenamefont {Limmer}, \citenamefont {Madden},\ and\
  \citenamefont {Chandler}}]{willard2013a}%
  \BibitemOpen
  \bibfield  {author} {\bibinfo {author} {\bibfnamefont {A.~P.}\ \bibnamefont
  {Willard}}, \bibinfo {author} {\bibfnamefont {D.~T.}\ \bibnamefont {Limmer}},
  \bibinfo {author} {\bibfnamefont {P.~A.}\ \bibnamefont {Madden}}, \ and\
  \bibinfo {author} {\bibfnamefont {D.}~\bibnamefont {Chandler}},\ }\bibfield
  {title} {\enquote {\bibinfo {title} {Characterizing heterogeneous dynamics at
  hydrated electrode surfaces},}\ }\href@noop {} {\bibfield  {journal}
  {\bibinfo  {journal} {J. Chem. Phys.}\ }\textbf {\bibinfo {volume} {138}},\
  \bibinfo {pages} {184702} (\bibinfo {year} {2013})}\BibitemShut {NoStop}%
\bibitem [{\citenamefont {Limmer}\ \emph {et~al.}(2015)\citenamefont {Limmer},
  \citenamefont {Willard}, \citenamefont {Madden},\ and\ \citenamefont
  {Chandler}}]{limmer2015b}%
  \BibitemOpen
  \bibfield  {author} {\bibinfo {author} {\bibfnamefont {D.~T.}\ \bibnamefont
  {Limmer}}, \bibinfo {author} {\bibfnamefont {A.~P.}\ \bibnamefont {Willard}},
  \bibinfo {author} {\bibfnamefont {P.~A.}\ \bibnamefont {Madden}}, \ and\
  \bibinfo {author} {\bibfnamefont {D.}~\bibnamefont {Chandler}},\ }\bibfield
  {title} {\enquote {\bibinfo {title} {Water {Exchange} at a {Hydrated}
  {Platinum} {Electrode} is {Rare} and {Collective}},}\ }\href {\doibase
  10.1021/acs.jpcc.5b08137} {\bibfield  {journal} {\bibinfo  {journal} {J.
  Phys. Chem. C}\ }\textbf {\bibinfo {volume} {119}},\ \bibinfo {pages}
  {24016--24024} (\bibinfo {year} {2015})}\BibitemShut {NoStop}%
\bibitem [{\citenamefont {Kattirtzi}, \citenamefont {Limmer},\ and\
  \citenamefont {Willard}(2017)}]{kattirtzi2017a}%
  \BibitemOpen
  \bibfield  {author} {\bibinfo {author} {\bibfnamefont {J.~A.}\ \bibnamefont
  {Kattirtzi}}, \bibinfo {author} {\bibfnamefont {D.~T.}\ \bibnamefont
  {Limmer}}, \ and\ \bibinfo {author} {\bibfnamefont {A.~P.}\ \bibnamefont
  {Willard}},\ }\bibfield  {title} {{\selectlanguage {english}\enquote
  {\bibinfo {title} {Microscopic dynamics of charge separation at the aqueous
  electrochemical interface},}\ }}\href {\doibase 10.1073/pnas.1700093114}
  {\bibfield  {journal} {\bibinfo  {journal} {Proc. Natl. Acad. Sci. U.S.A.}\
  }\textbf {\bibinfo {volume} {114}},\ \bibinfo {pages} {13374--13379}
  (\bibinfo {year} {2017})}\BibitemShut {NoStop}%
\bibitem [{\citenamefont {Merlet}\ \emph {et~al.}(2014)\citenamefont {Merlet},
  \citenamefont {Limmer}, \citenamefont {Salanne}, \citenamefont {{van Roij}},
  \citenamefont {Madden}, \citenamefont {Chandler},\ and\ \citenamefont
  {Rotenberg}}]{merlet2014a}%
  \BibitemOpen
  \bibfield  {author} {\bibinfo {author} {\bibfnamefont {C.}~\bibnamefont
  {Merlet}}, \bibinfo {author} {\bibfnamefont {D.~T.}\ \bibnamefont {Limmer}},
  \bibinfo {author} {\bibfnamefont {M.}~\bibnamefont {Salanne}}, \bibinfo
  {author} {\bibfnamefont {R.}~\bibnamefont {{van Roij}}}, \bibinfo {author}
  {\bibfnamefont {P.~A.}\ \bibnamefont {Madden}}, \bibinfo {author}
  {\bibfnamefont {D.}~\bibnamefont {Chandler}}, \ and\ \bibinfo {author}
  {\bibfnamefont {B.}~\bibnamefont {Rotenberg}},\ }\bibfield  {title} {\enquote
  {\bibinfo {title} {The electric double layer has a life of its own},}\
  }\href@noop {} {\bibfield  {journal} {\bibinfo  {journal} {J. Phys. Chem. C}\
  }\textbf {\bibinfo {volume} {118}},\ \bibinfo {pages} {18291--18298}
  (\bibinfo {year} {2014})}\BibitemShut {NoStop}%
\bibitem [{\citenamefont {Rotenberg}\ and\ \citenamefont
  {Salanne}(2015)}]{rotenberg2015a}%
  \BibitemOpen
  \bibfield  {author} {\bibinfo {author} {\bibfnamefont {B.}~\bibnamefont
  {Rotenberg}}\ and\ \bibinfo {author} {\bibfnamefont {M.}~\bibnamefont
  {Salanne}},\ }\bibfield  {title} {\enquote {\bibinfo {title} {Structural
  transitions at ionic liquid interfaces},}\ }\href@noop {} {\bibfield
  {journal} {\bibinfo  {journal} {J. Phys. Chem. Lett.}\ }\textbf {\bibinfo
  {volume} {6}},\ \bibinfo {pages} {4978--4985} (\bibinfo {year}
  {2015})}\BibitemShut {NoStop}%
\bibitem [{\citenamefont {Fedorov}\ and\ \citenamefont
  {Kornyshev}(2014)}]{fedorov2014a}%
  \BibitemOpen
  \bibfield  {author} {\bibinfo {author} {\bibfnamefont {M.~V.}\ \bibnamefont
  {Fedorov}}\ and\ \bibinfo {author} {\bibfnamefont {A.~A.}\ \bibnamefont
  {Kornyshev}},\ }\bibfield  {title} {\enquote {\bibinfo {title} {Ionic liquids
  at electrified interfaces},}\ }\href@noop {} {\bibfield  {journal} {\bibinfo
  {journal} {Chem. Rev.}\ }\textbf {\bibinfo {volume} {114}},\ \bibinfo {pages}
  {2978---3036} (\bibinfo {year} {2014})}\BibitemShut {NoStop}%
\bibitem [{\citenamefont {Scalfi}, \citenamefont {Salanne},\ and\ \citenamefont
  {Rotenberg}(2021)}]{scalfi_molecular_2021}%
  \BibitemOpen
  \bibfield  {author} {\bibinfo {author} {\bibfnamefont {L.}~\bibnamefont
  {Scalfi}}, \bibinfo {author} {\bibfnamefont {M.}~\bibnamefont {Salanne}}, \
  and\ \bibinfo {author} {\bibfnamefont {B.}~\bibnamefont {Rotenberg}},\
  }\bibfield  {title} {\enquote {\bibinfo {title} {Molecular {Simulation} of
  {Electrode}-{Solution} {Interfaces}},}\ }\href {\doibase
  10.1146/annurev-physchem-090519-024042} {\bibfield  {journal} {\bibinfo
  {journal} {Annual Review of Physical Chemistry}\ }\textbf {\bibinfo {volume}
  {72}},\ \bibinfo {pages} {189} (\bibinfo {year} {2021})}\BibitemShut
  {NoStop}%
\bibitem [{\citenamefont {Son}\ and\ \citenamefont
  {Wang}(2021)}]{son_image-charge_2021}%
  \BibitemOpen
  \bibfield  {author} {\bibinfo {author} {\bibfnamefont {C.~Y.}\ \bibnamefont
  {Son}}\ and\ \bibinfo {author} {\bibfnamefont {Z.-G.}\ \bibnamefont {Wang}},\
  }\bibfield  {title} {\enquote {\bibinfo {title} {Image-charge effects on ion
  adsorption near aqueous interfaces},}\ }\href {\doibase
  10.1073/pnas.2020615118} {\bibfield  {journal} {\bibinfo  {journal}
  {Proceedings of the National Academy of Sciences}\ }\textbf {\bibinfo
  {volume} {118}},\ \bibinfo {pages} {e2020615118} (\bibinfo {year}
  {2021})}\BibitemShut {NoStop}%
\bibitem [{\citenamefont {Merlet}\ \emph
  {et~al.}(2013{\natexlab{b}})\citenamefont {Merlet}, \citenamefont {P\'ean},
  \citenamefont {Rotenberg}, \citenamefont {Madden}, \citenamefont {Simon},\
  and\ \citenamefont {Salanne}}]{merlet2013b}%
  \BibitemOpen
  \bibfield  {author} {\bibinfo {author} {\bibfnamefont {C.}~\bibnamefont
  {Merlet}}, \bibinfo {author} {\bibfnamefont {C.}~\bibnamefont {P\'ean}},
  \bibinfo {author} {\bibfnamefont {B.}~\bibnamefont {Rotenberg}}, \bibinfo
  {author} {\bibfnamefont {P.~A.}\ \bibnamefont {Madden}}, \bibinfo {author}
  {\bibfnamefont {P.}~\bibnamefont {Simon}}, \ and\ \bibinfo {author}
  {\bibfnamefont {M.}~\bibnamefont {Salanne}},\ }\bibfield  {title} {\enquote
  {\bibinfo {title} {Simulating supercapacitors: Can we model electrodes as
  constant charge surfaces?}}\ }\href@noop {} {\bibfield  {journal} {\bibinfo
  {journal} {J. Phys. Chem. Lett.}\ }\textbf {\bibinfo {volume} {4}},\ \bibinfo
  {pages} {264--268} (\bibinfo {year} {2013}{\natexlab{b}})}\BibitemShut
  {NoStop}%
\bibitem [{\citenamefont {Ntim}\ and\ \citenamefont
  {Sulpizi}(2020)}]{ntim_role_2020}%
  \BibitemOpen
  \bibfield  {author} {\bibinfo {author} {\bibfnamefont {S.}~\bibnamefont
  {Ntim}}\ and\ \bibinfo {author} {\bibfnamefont {M.}~\bibnamefont {Sulpizi}},\
  }\bibfield  {title} {{\selectlanguage {english}\enquote {\bibinfo {title}
  {Role of image charges in ionic liquid confined between metallic
  interfaces},}\ }}\href {\doibase 10.1039/D0CP00409J} {\bibfield  {journal}
  {\bibinfo  {journal} {Phys. Chem. Chem. Phys.}\ }\textbf {\bibinfo {volume}
  {22}},\ \bibinfo {pages} {10786--10791} (\bibinfo {year} {2020})}\BibitemShut
  {NoStop}%
\bibitem [{\citenamefont {Scalfi}\ \emph
  {et~al.}(2020{\natexlab{b}})\citenamefont {Scalfi}, \citenamefont {Dufils},
  \citenamefont {Reeves}, \citenamefont {Rotenberg},\ and\ \citenamefont
  {Salanne}}]{scalfi_semiclassical_2020}%
  \BibitemOpen
  \bibfield  {author} {\bibinfo {author} {\bibfnamefont {L.}~\bibnamefont
  {Scalfi}}, \bibinfo {author} {\bibfnamefont {T.}~\bibnamefont {Dufils}},
  \bibinfo {author} {\bibfnamefont {K.~G.}\ \bibnamefont {Reeves}}, \bibinfo
  {author} {\bibfnamefont {B.}~\bibnamefont {Rotenberg}}, \ and\ \bibinfo
  {author} {\bibfnamefont {M.}~\bibnamefont {Salanne}},\ }\bibfield  {title}
  {\enquote {\bibinfo {title} {A semiclassical {Thomas}–{Fermi} model to tune
  the metallicity of electrodes in molecular simulations},}\ }\href {\doibase
  10.1063/5.0028232} {\bibfield  {journal} {\bibinfo  {journal} {The Journal of
  Chemical Physics}\ }\textbf {\bibinfo {volume} {153}},\ \bibinfo {pages}
  {174704} (\bibinfo {year} {2020}{\natexlab{b}})}\BibitemShut {NoStop}%
\bibitem [{\citenamefont {Schlaich}\ \emph {et~al.}(2020)\citenamefont
  {Schlaich}, \citenamefont {Jin}, \citenamefont {Bocquet},\ and\ \citenamefont
  {Coasne}}]{schlaich2020arXiv}%
  \BibitemOpen
  \bibfield  {author} {\bibinfo {author} {\bibfnamefont {A.}~\bibnamefont
  {Schlaich}}, \bibinfo {author} {\bibfnamefont {D.}~\bibnamefont {Jin}},
  \bibinfo {author} {\bibfnamefont {L.}~\bibnamefont {Bocquet}}, \ and\
  \bibinfo {author} {\bibfnamefont {B.}~\bibnamefont {Coasne}},\ }\href@noop {}
  {\enquote {\bibinfo {title} {Wetting transition of ionic liquids at metal
  surfaces: A computational approach to electronic screening using a virtual
  {Thomas}--{Fermi} fluid},}\ } (\bibinfo {year} {2020}),\ \bibinfo {note}
  {\url{https://arxiv.org/abs/2002.11526}},\ \Eprint
  {http://arxiv.org/abs/2002.11526} {arXiv:2002.11526 [physics.chem-ph]}
  \BibitemShut {NoStop}%
\bibitem [{\citenamefont {Gingrich}\ and\ \citenamefont
  {Wilson}(2010)}]{gingrich_ewald_2010}%
  \BibitemOpen
  \bibfield  {author} {\bibinfo {author} {\bibfnamefont {T.~R.}\ \bibnamefont
  {Gingrich}}\ and\ \bibinfo {author} {\bibfnamefont {M.}~\bibnamefont
  {Wilson}},\ }\bibfield  {title} {{\selectlanguage {english}\enquote {\bibinfo
  {title} {On the {Ewald} summation of {Gaussian} charges for the simulation of
  metallic surfaces},}\ }}\href {\doibase 10.1016/j.cplett.2010.10.010}
  {\bibfield  {journal} {\bibinfo  {journal} {Chem. Phys. Lett.}\ }\textbf
  {\bibinfo {volume} {500}},\ \bibinfo {pages} {178--183} (\bibinfo {year}
  {2010})}\BibitemShut {NoStop}%
\bibitem [{\citenamefont {Xie}\ \emph {et~al.}(2020)\citenamefont {Xie},
  \citenamefont {Fu}, \citenamefont {Niehaus},\ and\ \citenamefont
  {Joly}}]{xie_liquid-solid_2020}%
  \BibitemOpen
  \bibfield  {author} {\bibinfo {author} {\bibfnamefont {Y.}~\bibnamefont
  {Xie}}, \bibinfo {author} {\bibfnamefont {L.}~\bibnamefont {Fu}}, \bibinfo
  {author} {\bibfnamefont {T.}~\bibnamefont {Niehaus}}, \ and\ \bibinfo
  {author} {\bibfnamefont {L.}~\bibnamefont {Joly}},\ }\bibfield  {title}
  {\enquote {\bibinfo {title} {Liquid-{Solid} {Slip} on {Charged} {Walls}:
  {The} {Dramatic} {Impact} of {Charge} {Distribution}},}\ }\href {\doibase
  10.1103/PhysRevLett.125.014501} {\bibfield  {journal} {\bibinfo  {journal}
  {Physical Review Letters}\ }\textbf {\bibinfo {volume} {125}},\ \bibinfo
  {pages} {014501} (\bibinfo {year} {2020})}\BibitemShut {NoStop}%
\bibitem [{\citenamefont {Marin-Lafl\`eche}\ \emph {et~al.}(2020)\citenamefont
  {Marin-Lafl\`eche}, \citenamefont {Haefele}, \citenamefont {Scalfi},
  \citenamefont {Coretti}, \citenamefont {Dufils}, \citenamefont {Jeanmairet},
  \citenamefont {Reed}, \citenamefont {Serva}, \citenamefont {Berthin},
  \citenamefont {Bacon}, \citenamefont {Bonella}, \citenamefont {Rotenberg},
  \citenamefont {Madden},\ and\ \citenamefont
  {Salanne}}]{marin-lafleche_metalwalls_2020}%
  \BibitemOpen
  \bibfield  {author} {\bibinfo {author} {\bibfnamefont {A.}~\bibnamefont
  {Marin-Lafl\`eche}}, \bibinfo {author} {\bibfnamefont {M.}~\bibnamefont
  {Haefele}}, \bibinfo {author} {\bibfnamefont {L.}~\bibnamefont {Scalfi}},
  \bibinfo {author} {\bibfnamefont {A.}~\bibnamefont {Coretti}}, \bibinfo
  {author} {\bibfnamefont {T.}~\bibnamefont {Dufils}}, \bibinfo {author}
  {\bibfnamefont {G.}~\bibnamefont {Jeanmairet}}, \bibinfo {author}
  {\bibfnamefont {S.~K.}\ \bibnamefont {Reed}}, \bibinfo {author}
  {\bibfnamefont {A.}~\bibnamefont {Serva}}, \bibinfo {author} {\bibfnamefont
  {R.}~\bibnamefont {Berthin}}, \bibinfo {author} {\bibfnamefont
  {C.}~\bibnamefont {Bacon}}, \bibinfo {author} {\bibfnamefont
  {S.}~\bibnamefont {Bonella}}, \bibinfo {author} {\bibfnamefont
  {B.}~\bibnamefont {Rotenberg}}, \bibinfo {author} {\bibfnamefont {P.~A.}\
  \bibnamefont {Madden}}, \ and\ \bibinfo {author} {\bibfnamefont
  {M.}~\bibnamefont {Salanne}},\ }\bibfield  {title} {\enquote {\bibinfo
  {title} {{MetalWalls}: {A} classical molecular dynamics software dedicated to
  the simulation of electrochemical systems},}\ }\href {\doibase
  10.21105/joss.02373} {\bibfield  {journal} {\bibinfo  {journal} {Journal of
  Open Source Software}\ }\textbf {\bibinfo {volume} {5}},\ \bibinfo {pages}
  {2373} (\bibinfo {year} {2020})}\BibitemShut {NoStop}%
\bibitem [{\citenamefont {Berendsen}, \citenamefont {Grigera},\ and\
  \citenamefont {Straatsma}(1987)}]{berendsen1987a}%
  \BibitemOpen
  \bibfield  {author} {\bibinfo {author} {\bibfnamefont {H.~J.~C.}\
  \bibnamefont {Berendsen}}, \bibinfo {author} {\bibfnamefont {J.~R.}\
  \bibnamefont {Grigera}}, \ and\ \bibinfo {author} {\bibfnamefont {T.~P.}\
  \bibnamefont {Straatsma}},\ }\bibfield  {title} {\enquote {\bibinfo {title}
  {{The Missing Term in Effective Pair Potentials}},}\ }\href@noop {}
  {\bibfield  {journal} {\bibinfo  {journal} {J. Phys. Chem.}\ }\textbf
  {\bibinfo {volume} {91}},\ \bibinfo {pages} {6269--6271} (\bibinfo {year}
  {1987})}\BibitemShut {NoStop}%
\bibitem [{\citenamefont {Dang}(1995)}]{dang_mechanism_1995}%
  \BibitemOpen
  \bibfield  {author} {\bibinfo {author} {\bibfnamefont {L.~X.}\ \bibnamefont
  {Dang}},\ }\bibfield  {title} {\enquote {\bibinfo {title} {Mechanism and
  {Thermodynamics} of {Ion} {Selectivity} in {Aqueous} {Solutions} of
  18-{Crown}-6 {Ether}: {A} {Molecular} {Dynamics} {Study}},}\ }\href {\doibase
  10.1021/ja00131a018} {\bibfield  {journal} {\bibinfo  {journal} {J. Am. Chem.
  Soc.}\ }\textbf {\bibinfo {volume} {117}},\ \bibinfo {pages} {6954--6960}
  (\bibinfo {year} {1995})}\BibitemShut {NoStop}%
\bibitem [{\citenamefont {Werder}\ \emph {et~al.}(2003)\citenamefont {Werder},
  \citenamefont {Walther}, \citenamefont {Jaffe}, \citenamefont {Halicioglu},\
  and\ \citenamefont {Koumoutsakos}}]{werder_watercarbon_2003}%
  \BibitemOpen
  \bibfield  {author} {\bibinfo {author} {\bibfnamefont {T.}~\bibnamefont
  {Werder}}, \bibinfo {author} {\bibfnamefont {J.~H.}\ \bibnamefont {Walther}},
  \bibinfo {author} {\bibfnamefont {R.~L.}\ \bibnamefont {Jaffe}}, \bibinfo
  {author} {\bibfnamefont {T.}~\bibnamefont {Halicioglu}}, \ and\ \bibinfo
  {author} {\bibfnamefont {P.}~\bibnamefont {Koumoutsakos}},\ }\bibfield
  {title} {\enquote {\bibinfo {title} {On the {Water}-{Carbon} {Interaction}
  for {Use} in {Molecular} {Dynamics} {Simulations} of {Graphite} and {Carbon}
  {Nanotubes}},}\ }\href {\doibase 10.1021/jp0268112} {\bibfield  {journal}
  {\bibinfo  {journal} {J. Phys. Chem. B}\ }\textbf {\bibinfo {volume} {107}},\
  \bibinfo {pages} {1345--1352} (\bibinfo {year} {2003})}\BibitemShut {NoStop}%
\bibitem [{\citenamefont {Berg}, \citenamefont {Peter},\ and\ \citenamefont
  {Johnston}(2017)}]{berg_evaluation_2017}%
  \BibitemOpen
  \bibfield  {author} {\bibinfo {author} {\bibfnamefont {A.}~\bibnamefont
  {Berg}}, \bibinfo {author} {\bibfnamefont {C.}~\bibnamefont {Peter}}, \ and\
  \bibinfo {author} {\bibfnamefont {K.}~\bibnamefont {Johnston}},\ }\bibfield
  {title} {\enquote {\bibinfo {title} {Evaluation and {Optimization} of
  {Interface} {Force} {Fields} for {Water} on {Gold} {Surfaces}},}\ }\href
  {\doibase 10.1021/acs.jctc.7b00612} {\bibfield  {journal} {\bibinfo
  {journal} {Journal of Chemical Theory and Computation}\ }\textbf {\bibinfo
  {volume} {13}},\ \bibinfo {pages} {5610--5623} (\bibinfo {year}
  {2017})}\BibitemShut {NoStop}%
\end{thebibliography}
\end{document}